\documentclass[12pt]{article}
\pdfoutput=1
\usepackage{graphicx,amsfonts,amsthm,amsmath,amssymb,hyperref,color,yfonts,cite}
\newcommand{\Tr}{{\rm Tr}}
\newcommand{\bea}{\begin{eqnarray}\displaystyle}
\newcommand{\eea}{\end{eqnarray}}

\begin{document}
\makeatletter
\@addtoreset{equation}{section}
\makeatother
\renewcommand{\theequation}{\thesection.\arabic{equation}}
\vspace{1.8truecm}
%%%%%%%%%%%%%%%%%

{\LARGE{ \centerline{\bf Reconstructing the spacetime dual to a free matrix}}}  

\vskip.5cm 

\thispagestyle{empty} 
\centerline{{\large\bf Robert de Mello Koch$^{a,b,d}$\footnote{{\tt robert@zjhu.edu.cn}}, Pratik Roy$^{b,d}$\footnote{{\tt roy.pratik92@gmail.com}} and
 Hendrik J.R. Van Zyl a$^{c,d}$\footnote{\tt hjrvanzyl@gmail.com}}}
\vspace{.8cm}
\centerline{{\it $^{a}$School of Science, Huzhou University, Huzhou 313000, China,}}
\vspace{.8cm}
\centerline{{\it $^{b}$School of Physics and Mandelstam Institute for Theoretical Physics,}}
\centerline{{\it University of the Witwatersrand, Wits, 2050, }}
\centerline{{\it South Africa }}
\vspace{.8cm}
\centerline{{\it $^{c}$The Laboratory for Quantum Gravity \& Strings,}}
\centerline{{\it Department of Mathematics \& Applied Mathematics,}}
\centerline{{\it University of Cape Town, Cape Town, South Africa}}
\vspace{.8cm}
\centerline{{\it $^{d}$ The National Institute for Theoretical and Computational Sciences,}} \centerline{{\it Private Bag X1, Matieland, South Africa}}

\vspace{1truecm}

%%%%%%%%%%%%%%%%%
\thispagestyle{empty}

\centerline{\bf ABSTRACT}

\vskip.2cm 
In this paper we consider the collective field theory description of the singlet sector of a free matrix field in 2+1 dimensions. This necessarily involves the study of $k$-local collective fields, which are functions of $2k+1$ coordinates. We argue that these coordinates have a natural interpretation: the $k$-local collective field is a field defined on an AdS$_4\times$S$^{k-2}\times$S$^{k-1}$ spacetime. The modes of a harmonic expansion on the S$^{k-2}\times$S$^{k-1}$ portion of the spacetime leads to the spinning bulk fields of the dual gravity theory.

\setcounter{page}{0}
\setcounter{tocdepth}{2}
\newpage
\tableofcontents
\setcounter{footnote}{0}
\linespread{1.1}
\parskip 4pt

{}~
{}~

\section{Introduction}

Collective field theory \cite{Jevicki:1979mb,Jevicki:1980zg} provides a constructive approach to gauge theory/gravity dualities \cite{Maldacena:1997re,Gubser:1998bc,Witten:1998qj}. By expressing the gauge theory in terms of invariant collective fields, the original loop expansion parameter of the field theory ($\hbar$) is replaced by ${1\over N}$ \cite{Jevicki:1979mb,Jevicki:1980zg}. The theory of the collective fields is the gravitational theory - see \cite{Das:1990kaa,Das:2003vw,deMelloKoch:2010wdf,Mintun:2014gua,deMelloKoch:2023ylr} for some compelling examples. 

The collective field theory description of the singlet sector of a free hermitian massless scalar matrix theory in 2+1 dimensions was initiated in \cite{deMelloKoch:2024ewt}. The action of the theory is
\bea
S=\int d^3 x\,{1\over 2}\,\Tr (\partial_\mu\phi\,\partial^\mu\phi)
\eea
The free matrix is expected to be dual to the tensionless limit of a string theory \cite{Mikhailov:2002bp}. In this setting, the collective fields are $k$-local operators obtained by tracing a product of $k$-matrices. Following \cite{deMelloKoch:2010wdf} we work in an equal $x^+$ description so the $k$-local operator 
\bea
\sigma_k(x^+,x_1^-,x_1,x_2^-,x_2,\cdots,x_k^-,x_k)&=&\Tr\Big(\phi(x^+,x_1^-,x_1)\phi(x^+,x_2^-,x_2)\cdots\phi(x^+,x_k^-,x_k)\Big)\cr
&&
\eea
is a function of $2k+1$ coordinates. This represents a challenge to the collective field theory construction of holography: since the gravitational theory is defined in an asymtotically AdS$_4$ spacetime, it is clear that 4 coordinates are needed to describe the dual spacetime. What is the interpretation of the remaining $2k-3$ coordinates? If the collective field theory does indeed reconstruct the gravity theory, these additional coordinates must also have a natural gravity interpretation. This is a remarkable claim. It is precisely this question that is considered in this paper.

To construct the mapping between the conformal field theory and the dual gravity theory\footnote{We use little letters to refer to the coordinates of spacetime of the conformal field theory and capital letters to refer to the coordinates of AdS$_4$.}, it is helpful to perform a Fourier transform on both sides of the duality, replacing $x^-$ by $p^+$ in the conformal field theory, and $X^-$ by $P^+$ in AdS$_4$. In terms of these conformal field theory coordinates, the coordinates of the bulk AdS$_4$ spacetime are \cite{deMelloKoch:2024ewt,deMelloKoch:2024juz}
\bea
X&=&{\sum_{i=1}^k p_i^+ x_i\over \sum_{j=1}^k p_j^+}
\qquad
Z\,\,=\,\,{\sqrt{\sum_{i=1}^k p_i^+ v_i^2}\over \left(\sum_{j=1}^k p_j^+\right)^{3\over 2}}\qquad P^+\,\,=\,\,\sum_{i=1}^k p_i^+\qquad X^+=x^+
\label{coordmap}
\eea
where
\bea
v_i&=&\sum_{j=1}^k p_j^+(x_i-x_j)\label{defv}
\eea
To understand the role of the additional $2k-3$ coordinates, note that the $k$-local collective field is a product of scalars, each of which transforms in the dimension $\Delta={1\over 2}$ and spin $s=0$ representation\footnote{We label irreducible representations of SO(2,3) using the dimension and spin as $(\Delta,s)$.} of SO(2,3). The collective field is therefore in a highly reducible representation, given by taking the tensor product of $k$ copies of the $({1\over 2},0)$ representation with cyclic symmetry (from the cyclicity of the trace) imposed. The reducible representation of a given collective field decomposes into a direct sum of an infinite number of irreducible representations. Each irreducible representation corresponds to a unique primary operator and the collective field packages this infinite collection of primaries. For a detailed explanation of what primaries can be recovered from the $k$-local collective field the reader can consult \cite{deMelloKoch:2024ewt}. The extra coordinates present in the collective field theory description are needed to collect these primary operators in a systematic way.
  
Each primary operator contained in the collective field is dual to a bulk field. The beautiful series of papers \cite{Metsaev:1999ui,Metsaev:2003cu,Metsaev:2005ws,Metsaev:2013kaa,Metsaev:2015rda,Metsaev:2022ndg} has developed the description of these spinning fields in light front AdS. In particular, each field corresponds to a representation $(\Delta,s)$ of SO(2,3) and the bulk equation of motion for each such field is known. Using the relation between the coordinates (\ref{coordmap}), this infinite set of bulk equations of motion can be translated into equations in the collective field theory. In the collective field theory they become equations which set the second quadratic Casimir of SO(2,3) to the correct value for the $(\Delta,s)$ representation, i.e. they extract the correct primary operator from the multilocal collective field \cite{deMelloKoch:2024ewt}. This demonstrates that the infinite number of bulk equations of motion are obeyed in the collective field theory description. This is almost a solution to the bulk reconstruction problem for the complete set of excitations of this tensionless string theory. All that is missing is a demonstration that the reconstructed bulk field obeys the correct boundary condition, i.e. that it obeys the so called GKPW rule \cite{Gubser:1998bc,Witten:1998qj}. This is the problem we address in this article. We will see that it is intimately related to the interpretation of the additional $2k-3$ coordinates of the collective field theory description.

Our intuition comes from the bilocal collective field which has $k=2$. In this case, the problem we consider has already been solved \cite{deMelloKoch:2010wdf}. For the bilocal 
\bea
\sigma_2(x^+,x_1^-,x_1,x_2^-,x_2)&=&\Tr \Big(\phi(x^+,x_1^-,x_1)\phi(x^+,x_2^-,x_2)\Big)
\eea
where $2k-3=1$ there is a single extra coordinate and it is an angle. This bilocal field can be written as the sum of a large $N$ expectation value of the field, denoted $\sigma_2^0(x^+,x^-_1,x_1,x^-_2,x_2)$, plus a fluctuation $\eta_2(x^+,x^-_1,x_1,x^-_2,x_2)$ as follows\footnote{The coefficient of $\eta_2$ ensures that it has an order 1 two point function as we take $N\to\infty$. Here $\eta_2$ is the normal ordered trace of a product of two matrices.}
\bea
\sigma_2(x^+,x^-_1,x_1,x^-_2,x_2)&=&\sigma_2^0(x^+,x^-_1,x_1,x^-_2,x_2)
+{1\over N}\eta_2(x^+,x^-_1,x_1,x^-_2,x_2)
\eea
It is the fluctuation $\eta_2(x^+,x^-_1,x_1,x^-_2,x_2)$ that is identified with the dynamical higher spin gravity fields. The mapping between the coordinates of the collective field theory and those of the bulk gravity is given by
\bea
X&=& \frac{p_1^+ x_1+p_2^+ x_2}{p_1^++p_2^+}\qquad
Z\,\,=\,\,\frac{\sqrt{p_1^+ p_2^+} (x_1-x_2)}{p_1^++p_2^+}\qquad X^+=x^+\cr
&&\cr
&&\cr
P^+&=& p_1^++p_2^+\qquad\qquad
\theta\,\,=\,\,2 \tan ^{-1}\left(\sqrt{\frac{p_2^+}{p_1^+}}\right)
\label{bicoordmap}
\eea
A Fourier expansion of the bilocal collective field with respect to this angle gives the different bulk fields as the Fourier modes
\bea
\Phi&=&\sum_{s=0}^\infty
\left(\cos (2s\theta) {A^{XX\cdots XX}(X^+,P^+,X,Z)\over Z}+\sin (2s\theta) {A^{XX\cdots XZ}(X^+,P^+,X,Z)\over Z}\right)\cr\cr
 &=&2\pi P^+\sin\theta\,\,  
\eta_2(X^+,P^+\cos^2{\theta\over 2},X+Z \tan{\theta\over 2},P^+\sin^2{\theta\over 2},X-Z\cot {\theta \over 2})\label{fldident}
\eea
Here $A^{XX\cdots XX}(X^+,P^+,X,Z)$ and $A^{XX\cdots XZ}(X^+,P^+,X,Z)$ are the two independent and physical components of the spin $2s$ gauge field in light cone gauge \cite{Metsaev:1999ui}. If one takes the limit that $Z\to 0$, using the above identification of the bulk field in terms of the fluctuation $\eta$, one finds a close connection to all of the conserved spinning currents (for $s>0$) and to the scalar field $\Tr (\phi^2)$. This is the complete set of primary operators of the conformal field theory packaged in the bilocal collective field. The basis of this connection comes from the identity obtained by evaluating $\cos(2s\theta)$ with the angle defined in (\ref{coordmap})
\bea
&&(p_1^++p_2^+)^{2s}\, \cos \left(2s\theta_1\right)\,\,=\,\,(p_1^++p_2^+)^{2s}\, \cos \left(4s \tan^{-1}\sqrt{p_2^+\over p_1^+}\right)\cr\cr
&&\qquad\qquad\,\,=\,\,(2 s)! (4 s-1)!! \sum _{k=0}^{2 s}\sum _{l=0}^{2 s-k} \frac{(-1)^k \,\, (p_1^+)^k (p_2^+)^{2s-k}}{k!(2s-k)!(2 k-1)!!(4s-2 k-1)!!}
\label{BiAngle}
\eea
There is an obvious connection to the formula for the primary current
\bea
{\cal O}_s^{++\cdots+}&=&(2 s)! (4 s-1)!! \sum _{k=0}^{2 s}\sum _{l=0}^{2 s-k} \frac{(-1)^k \,\,\partial^{+\,k}\phi\,\partial^{+\,2s-k}\phi}{k!(2s-k)!(2 k-1)!!(4s-2 k-1)!!}
\label{spinningcurrent}
\eea
This connection was used in \cite{Mintun:2014gua} to argue that the bilocal collective field theory correctly reproduces the GKPW rule, after changing from de Donder to light cone gauge. In this way, the change to collective bilocal fields and the subsequent change of coordinates (\ref{coordmap}) provides a detailed and exact identification between independent degrees of freedom in the conformal field theory and physical and independent degrees of freedom in the higher spin gravity \cite{deMelloKoch:2023ngh}. The collective field theory of the bilocals explicitly realizes \cite{deMelloKoch:2021cni} entanglement wedge reconstruction \cite{Dong:2016eik} and it is manifestly \cite{deMelloKoch:2022sul} consistent with the principle of the holography of information \cite{Laddha:2020kvp}. This discussion makes it clear that the extra coordinate $\theta$, in the case of the bilocal collective fields, is responsible for a systematic organization of the primary operators contained in the collective field.
  
It is natural to guess that for other values of $k$ something similar will happen and the extra coordinates are again responsible for a systematic accounting of the primaries captured in the collective fields. Exploring this expectation is the key goal of this article.

In Section \ref{Bcoords} we derive a set of $2k+1$ bulk coordinates for the $k$-local collective field. The AdS bulk coordinates given in (\ref{coordmap}) are 4 of these coordinates. The remaining $2k-3$ coordinates are determined by requiring that they are independent\footnote{The derivative of these $2k-3$ coordinates with respect to the bulk coordinates vanishes.} of these 4 bulk AdS coordinates. In Section \ref{Interp} we argue that the complete set of coordinates of the $k$-local collective field parametrizes the AdS$_4\times$S$^{k-2}\times$S$^{k-1}$ spacetime. The modes of a harmonic expansion on the S$^{k-2}\times$S$^{k-1}$ factor of the geometry are the bulk fields of the dual gravity theory. In Section \ref{BC}, we show in detail for the trilocal collective field, that the GKPW rule does indeed hold. Finally, we present our conclusions in Section \ref{conclusions}.

\section{Bulk coordinates}\label{Bcoords}

The mapping between the coordinates of the collective field theory and those of the dual AdS$_4$ spacetime are given in (\ref{coordmap}) above. Derivatives with respect to the AdS coordinates are given by
\bea
{\partial\over\partial X}&=&\sum_{j=1}^k {\partial\over\partial x_j}\qquad\qquad\qquad {\partial\over\partial Z}\,\,=\,\,{1\over Z}{\sum_{i=1}^k v_i {\partial\over\partial x_i}\over \sum_{j=1}^k p_j^+}\cr\cr
{\partial\over\partial P^+}&=&{\sum_{i=1}^k p_i^+{\partial\over\partial p_i^+}\over\sum_{j=1}^k p_j^+}\qquad\qquad
{\partial\over\partial X^+}={\partial\over\partial x^+}\label{derblkcoordinates}
\eea

\subsection{Trilocal Collective Field}

We start with a study of the trilocal collective field. For this case $2k-3=3$ so the collective field theory has 3 coordinates in addition to those associated with the AdS$_4$ bulk. With these three additional coordinates the mapping between the coordinates of the collective field theory and those of the bulk AdS theory exchanges
\bea
\{ x^+,x_1,x_2,x_3,p_1^+,p_2^+,p_3^+\}\qquad\leftrightarrow\qquad\{X^+,P^+,X,Z,u_1,u_2,u_3\}
\eea
Our first task is to determine suitable coordinates $u_1$, $u_2$ and $u_3$. First, we require that
\bea
{\partial\over\partial P^+}u_i &=&0\qquad i=1,2,3\label{constr1}
\eea
which restricts how the $u_i$ depend on $p_1^+,p_2^+,p_3^+$: they depend on these three coordinates through the two ratios ${p_1^+\over p_2^+}$ and ${p_1^+\over p_3^+}$. There is some arbitrariness in the choice of exactly which two ratios to take, but all choices are simply reparametrisations of each other. We also require that
\bea
{\partial\over\partial X}u_i &=&0\qquad i=1,2,3\label{constr2}
\eea
which restricts how the $u_i$ depend on $x_1,x_2,x_3$: they only depend on the two differences $x_{12}=x_1-x_2$ and $x_{13}=x_1-x_3$. There is again some arbitrariness in this choice. Other choices are again simply reparametrisations of this one. Finally, we must also require that
\bea
{\partial\over\partial Z}u_i &=&0\qquad i=1,2,3\label{constr3}
\eea
which restricts how the $u_i$ depend on the differences $x_{12}$ and $x_{13}$: they only depend on the ratio ${x_{12}\over x_{13}}$. The conclusion is that the three constraints (\ref{constr1}), (\ref{constr2}) and (\ref{constr3}) have constrained the dependence on the six coordinates $x_1,x_2,x_3,p_1^+,p_2^+,p_3^+$ to a dependence on the three coordinates
\bea
u_1&=&{x_{12}\over x_{13}},\qquad u_2\,\,=\,\,{p_1^+\over p_2^+}\qquad u_3\,\,=\,\,{p_1^+\over p_3^+}
\eea

To check this argument we can test whether the change of coordinates between $x_1,x_2,x_3,p_1^+,p_2^+,p_3^+$ and $X,Z,P^+,u_1,u_2,u_3$ is a valid change of coordinates i.e. if it is invertible. It is indeed invertible, with the result
\bea
X&=&{p_1^+x_1+p_2^+x_2+p_3^+x_3\over p_1^++p_2^++p_3^+}
\qquad\qquad\qquad
Z\,\,=\,\,{\sqrt{p_1^+ v_1^2+p_2^+ v_2^2+p_3^+ v_3^2}\over \left(p_1^++p_2^++p_3^+\right)^{3\over 2}}\cr\cr 
P^+&=&p_1^++p_2^++p_3^+\qquad\qquad\qquad\qquad\,\, X^+\,\,=\,\,x^+\cr\cr
 u_1&=&{x_{12}\over x_{13}}\qquad\qquad\qquad u_2\,\,=\,\,{p_1^+\over p_2^+}\qquad\qquad\qquad u_3\,\,=\,\,{p_1^+\over p_3^+}\label{coordmaptri}
\eea
where $v_i$ is defined in (\ref{defv}) and the inverse transformation is
\bea
x_1&=& X+Z\frac{u_1 u_3+u_2}{\sqrt{u_2u_3(u_1^2 u_3+u_1^2-2u_1+u_2+1)}}\qquad\qquad p_1^+\,\,=\,\, \frac{P^+ u_2 u_3}{u_2 u_3+u_2+u_3}\cr\cr
x_2&=& X-Z\frac{\sqrt{u_2}(u_1 u_3+u_1-1)}{\sqrt{u_3(u_1^2 u_3+u_1^2-2u_1+u_2+1)}}\qquad\qquad\quad
p_2^+\,\,=\,\, \frac{P^+ u_3}{u_2 u_3+u_2+u_3}\cr\cr
x_3&=& X+Z\frac{\sqrt{u_3}(u_1-u_2-1)}{\sqrt{u_2(u_1^2 u_3+u_1^2-2u_1+u_2+1)}}
\qquad\qquad\quad p_3^+\,\,=\,\,\frac{P^+ u_2}{u_2 u_3+u_2+u_3}
\eea

Given the connection between (\ref{BiAngle}) and (\ref{spinningcurrent}), the appearance of $u_2$ and $u_3$ as bulk coordinates is rather natural. These two variables are needed to produce a polynomial in the $p_i^+$ that reproduces the combination of derivatives needed to construct a primary operator, exactly as in (\ref{spinningcurrent}) for the bilocal case. The variable $u_1$ at first appears much less natural. What is the role of this coordinate? Recall that we make contact with primary operators in the conformal field theory in the $Z\to 0$ limit of the bulk theory. Since the light cone momenta $p_i^+$ are all positive, the formula for $Z$ in (\ref{coordmaptri}) implies that we must send each of the $v_i\to 0$. It is simple to verify that
\bea
x_i-x_j={v_i-v_j\over p_1^++p_2^++p_3^+}
\eea
Thus, sending all of the $v_i\to 0$ implies that we must take $x_1$, $x_2$ and $x_3$ to be coincident. There are many different ways we can do this. For example, we might first take $x_{12}\to 0$ by computing the OPE between the fields at $x_1$ and $x_2$. We could follow this by taking the OPE of the result of this first OPE with the field at $x_3$. Alternatively, we could have started with the OPE between the fields at $x_1$ and $x_3$, and as a second step take the OPE with the field at $x_2$. The OPE channel we use is dictated by the locations of the operators. The rule is that we must take the product of operators that are closest first. The specific linear combination of primaries we obtain depends on which OPE channel we use, and so on how we take the coincident limit. Lets focus on the case that $x_1>x_2$ and $x_1>x_3$, for example. In this case, to obtain the coincident limit we can set
\bea
x_{12}&=&\delta_A\epsilon\qquad\qquad x_{13}\,\,=\,\, \delta_B\epsilon\qquad\qquad u_1={\delta_A\over\delta_B}
\eea
and take $\epsilon\to 0$. If $0\le u_1<{1\over 2}$ and if the OPE is to converge, we must first take the OPE between the fields at $x_1$ and $x_2$, and then take the OPE with the field at $x_3$. If ${1\over 2}<u_1<1$ and if the OPE is to converge, we must first take the OPE between the fields at $x_2$ and $x_3$, and then take the OPE with the field at $x_1$. Thus, one role of the variable $u_1$ is to keep track of what OPE must be performed to take the coincident limit and extract local primary operators.

\subsection{Quadlocal Collective Field}

Next, consider the quadlocal collective field. In this case there are $2k-3=5$ coordinates in addition to the 4 coordinates of the AdS$_4$ bulk. The mapping between the coordinates of the collective field theory and those of the bulk AdS theory exchanges
\bea
\{ x^+,x_1,x_2,x_3,x_4,p_1^+,p_2^+,p_3^+,p_4^+\}\qquad\leftrightarrow\qquad\{X^+,P^+,X,Z,u_1,u_2,u_3,u_4,u_5\}
\eea
Working exactly as above we find
\bea
X&=&{p_1^+x_1+p_2^+x_2+p_3^+x_3+p_4^+x_4\over p_1^++p_2^++p_3^++p_4^+}
\qquad\qquad\qquad
Z\,\,=\,\,{\sqrt{p_1^+ v_1^2+p_2^+ v_2^2+p_3^+ v_3^2+p_4^+v_4^2}\over \left(p_1^++p_2^++p_3^++p_4^+\right)^{3\over 2}}\cr\cr 
P^+&=&p_1^++p_2^++p_3^++p_4^+\qquad\qquad\qquad\qquad\quad\,\, X^+\,\,=\,\,x^+\cr\cr
u_1&=&{x_{12}\over x_{14}},\qquad u_2\,\,=\,\,{x_{13}\over x_{14}}\qquad u_3\,\,=\,\,{p_1^+\over p_2^+}\qquad u_4\,\,=\,\,{p_1^+\over p_3^+}\qquad u_5\,\,=\,\,{p_1^+\over p_4^+}
\eea
and the inverse transformation
\bea
x_1&=& X-Z\frac{(u_1u_4 u_5+u_3 (u_2u_5+u_4))}{\sqrt{f(u_1,u_2,u_3,u_4,u_5)}}\cr\cr
x_2&=&X+Z\frac{u_3(u_5 (u_1-u_2)+u_4 (u_1u_5+u_1-1))}{\sqrt{f(u_1,u_2,u_3,u_4,u_5)}}\cr\cr
x_3&=&X+Z\frac{u_4(u_5 (u_2-u_1)+u_3 (u_2u_5+u_2-1))}{\sqrt{f(u_1,u_2,u_3,u_4,u_5)}}\cr\cr
x_4&=&X+Z\frac{u_5(-u_1u_4+u_3 (-u_2+u_4+1)+u_4)}{\sqrt{f(u_1,u_2,u_3,u_4,u_5)}}\cr\cr
p_1^+&=&\frac{u_3 u_4 u_5 P^+}{u_3 (u_4 u_5+u_4+u_5)+u_4 u_5}\cr\cr
p_2^+&=&\frac{u_4 u_5 P^+}{u_3 (u_4 u_5+u_4+u_5)+u_4 u_5}\cr\cr
p_3^+&=&\frac{u_3 u_5 P^+}{u_3 (u_4 u_5+u_4+u_5)+u_4 u_5}\cr\cr
p_4^+&=&\frac{u_3 u_4 P^+}{u_3 (u_4 u_5+u_4+u_5)+u_4 u_5}
\label{QuadLocalMap}
\eea
where
\bea
f(u_1,u_2,u_3,u_4,u_5)&=&u_3u_4u_5\Big(u_5(u_1-u_2)^2+u_1u_4 (u_1u_5+u_1-2)\cr\cr
&&+u_3\big(u_2(u_2u_5+u_2-2)+u_4+1\big)+u_4\Big)\label{defpolyf}
\eea

The interpretation of the extra coordinates $u_i$ is clear: one role of the pair of variables $u_1$ and $u_2$ is to track what OPE channel is used when taking $Z\to 0$, while the variables $u_3$, $u_4$ and $u_5$ are needed to build the polynomials in the $p_i^+$ that provide the correct combination of derivatives to produce the primaries.

\subsection{$k$-local Collective Field}

The results given above suggest a natural generalization. The $k$-local collective field has a total of $2k+1$ coordinates. The corresponding bulk coordinates are
\bea
P^+&=&\sum_{i=1}^k p_i^+\qquad
X\,\,=\,\,{\sum_{i=1}^k p_i^+x_i\over \sum_{j=1}^k p_j^+}
\qquad
Z\,\,=\,\,{\sqrt{\sum_{i=1}^kp_i^+ v_i^2}\over \left(\sum_{j=1}^k p_j^+\right)^{3\over 2}}\qquad X^+=x^+
\eea
as well as the additional $k-2$ coordinates of the form
\bea
u_j={x_1-x_{j+1}\over x_1-x_k}\qquad j=1,2,\cdots,k-2
\eea
and the $k-1$ coordinates of the form
\bea
u_{k-2+j}={p^+_1\over p^+_{j+1}}\qquad j=1,2,\cdots k-1
\eea
One role of the extra coordinates $u_i$ with $i=1,2,\cdots,k-2$ is to keep track of what OPE channel is used when taking $Z\to 0$, while the variables $u_{k-2+i}$ with $i=1,2,\cdots k-1$ are needed to build the polynomials in the $p_i^+$ that provide the combination of derivatives which produce the primaries.

Given this set of bulk coordinates, we can ask how they are to be interpreted. We turn to this question in the next section.

\section{Interpretation of the extra coordinates}\label{Interp}

In light cone gauge a complete gauge fixing of the gravity has been carried out in \cite{Metsaev:1999ui}. The massless fields (correponding to the bilocal collective field) of the higher spin gravity theory are a scalar field plus an infinite collection of spinning gauge fields. In light cone gauge the choice of gauge eliminates all $+$ polarizations of the spinning gauge fields, while solving the associated gauge constraints removes the $-$ polarizations. Thus, the completely gauge fixed description involves only the $X$ and $Z$ components of tensors. The spin part of rotations, denoted $M^{XZ}$, is related to the single angle $\theta$ that participates in the bilocal map as follows
\bea
M^{XZ}={\partial\over\partial\theta}
\eea
This provides a nice interpretation of the extra angle $\theta$ of the bilocal collective field and it suggests that a study of $M^{XZ}$ will be useful for the study of multilocal collective fields. 

\subsection{Trilocal Collective Field}\label{trilocalanalysis}

In what follows, we focus on the trilocal collective field. Notice that all of the primaries collected in the trilocal collective field are dual to massive fields in the AdS$_4$ gravity. In the case of the trilocal, it is possible to express $M^{XZ}$ in terms of the conformal field theory coordinates \cite{deMelloKoch:2024ewt}. The result is
\bea
M^{XZ}&=&{\sum_{i,j=1}^3 p_j^+(x_i-x_j)^2{\partial\over\partial x_i}-2\sum_{i=1}^3p_i^+v_i{\partial\over\partial p_i^+}\over 2\sqrt{\sum_{i>j=1}^3p_i^+p_j^+(x_i-x_j)^2}}
\eea
A tedious but otherwise straight forward computation shows that
\bea
M^{XZ}&=&\frac{u_2u_3+u_2+u_3}{\sqrt{u_2u_3}\sqrt{u_1^2+u_1^2 u_3-2 u_1+u_2+1}}\left(\frac{(1-u_1)}{2}u_1{\partial\over\partial u_1}-u_1 u_2{\partial\over\partial u_2}-u_3{\partial\over\partial u_3}\right)\cr
&&
\eea
The fact that $M^{XZ}$ is expressed entirely in terms of $u_1,u_2$ and $u_3$ is an encouraging sign.

At this point it is useful to recall some of the features of the light cone description of the gravity theory developed in \cite{Metsaev:1999ui,Metsaev:2003cu,Metsaev:2005ws,Metsaev:2013kaa,Metsaev:2015rda,Metsaev:2022ndg}. The equation of motion is given by
\bea
\left(2\partial^+\partial^-+\partial_X^2+\partial_Z^2-{A\over Z^2}\right)|\phi\rangle &=&0
\label{MEoM}
\eea
where $A$ is called the AdS mass operator. It can be written as 
\bea
A=\kappa^2-{1\over 4}
\eea
$M^{XZ}$, together with the operator $\kappa$ above and their commutator (which we call $\tau$) close an su(2) algebra. Bulk AdS$_4$ fields in a definite SU(2) representation, with a definite value for $\kappa$, are in a definite conformal representation and hence they correspond to a particular primary operator in the conformal field theory. This follows because, as we explain below, the SU(2) Casimir is related to the SO(2,3) Casimir. Since we expect that the extra coordinates $u_i$ are organizing primary operators, it is a useful exercise to study this su(2) algebra. Towards this end note that the operator $\kappa$, written in terms of the coordinates of the collective field, is \cite{deMelloKoch:2024ewt}
\bea
\kappa&=&\sqrt{p_1^+p_2^+p_3^+\over p_1^++p_2^++p_3^+}\left({x_2-x_3\over p_1^+}{\partial\over\partial x_1}+{x_3-x_1\over p_2^+}{\partial\over\partial x_2}+{x_1-x_2\over p_3^+}{\partial\over\partial x_3}\right)
\eea 
$\kappa$ can also be expressed in terms of $u_1$, $u_2$ and $u_3$ as follows
\bea
\kappa&=&\frac{u_1^2 (u_3+1)-2 u_1+u_2+1}{\sqrt{u_2 u_3+u_2+u_3}}\,{\partial\over\partial u_1}
\eea 
The operator $\tau$ in terms of the coordinates of the collective field is
\bea
\tau&=&\frac{\sqrt{p^+_1 p^+_2 p^+_3}}{\sqrt{p^+_1+p^+_2+p^+_3} \sqrt{p^+_1 p^+_2 (x_1-x_2)^2+p^+_1 p^+_3 (x_1-x_3)^2+p^+_2 p^+_3 (x_2-x_3)^2}}\times\cr\cr
 &&\Bigg((p^+_1+p^+_2+p^+_3) \left((x_2-x_3) \frac{\partial}{\partial p^+_1}+(x_3-x_1) \frac{\partial}{\partial p^+_2}+(x_1-x_2) \frac{\partial}{\partial p^+_3}\right)\cr\cr
&&-\frac{v_1 (x_2-x_3) }{2 p^+_1}\frac{\partial}{\partial x_1}-\frac{v_2 (x_3-x_1) }{2 p^+_2}\frac{\partial}{\partial x_2}-\frac{v_3 (x_1-x_2)}{2 p^+_3}\frac{\partial}{\partial x_3}\Bigg)
\eea
It also has an expression in terms of $u_1$, $u_2$ and $u_3$, but since the formula is not enlightening and we do not need it for any computations below, we will not quote it. The three generators
\bea
G_1&=&-\kappa\qquad G_2\,\,=\,\,2\tau\qquad G_3\,\,=\,\,-2M^{XZ}\label{GDefnd}
\eea
close the usual su(2) algebra
\bea
[G_1,G_2]&=&G_3\qquad [G_2,G_3]\,\,=\,\,G_1\qquad [G_3,G_1]\,\,=\,\,G_2
\eea
It is useful to change from $u_1$, $u_2$ and $u_3$ to the three angles $\alpha_1$, $\alpha_2$ and $\alpha_3$
\bea
\tan(\alpha_1)&=&\frac{1-u_1-u_1 u_3}{\sqrt{u_2 u_3+u_2+u_3}}\cr\cr
\tan(\alpha_2)&=&-\frac{\sqrt{u_3+1} \sqrt{u_2 u_3}}{u_3}\cr\cr
\tan(\alpha_3)&=&\sqrt{u_3}
\eea
of a three sphere S$^3=$S$^1\times$S$^2$. The inverse coordinate transformation is
\bea
u_1&=&\cos ^2(\alpha_3) \left(1-\tan (\alpha_1)\sec(\alpha_2)\tan(\alpha_3)\right)\cr\cr
u_2&=&\tan ^2(\alpha_2) \sin ^2(\alpha_3)\cr\cr
u_3&=&\tan ^2(\alpha_3)
\eea
In terms of these angles, the mapping between bulk and boundary becomes
\bea
x_1&=& X-Z\frac{\cos(\alpha_1)\cot (\alpha_3)-\cos(\alpha_2)\sin (\alpha_1)}{\sin (\alpha_2)}\qquad\qquad p_1^+\,\,=\,\, P^+ \sin ^2(\alpha_2) \sin ^2(\alpha_3)\cr\cr
x_2&=& X-Z\sin (\alpha_1)\tan(\alpha_2)\qquad\qquad\qquad\qquad\qquad\qquad
p_2^+\,\,=\,\, P^+ \cos ^2(\alpha_2)\cr\cr
x_3&=& X-Z{\cos (\alpha_1)\tan (\alpha_3)+\cos (\alpha_2)\sin (\alpha_1)\over\sin (\alpha_2)}
\qquad\qquad p_3^+\,\,=\,\,P^+ \sin ^2(\alpha_2) \cos ^2(\alpha_3)\cr
&&\label{trliccoc}
\eea

A significant advantage of this change of variables is that it takes the generators defined in (\ref{GDefnd}) to the standard form of the su(2) Killing vectors on $S^3$, which are
\bea
G_1&=&{\partial\over\partial\alpha_1}\cr\cr
G_2&=&-\sin\alpha_1\,\cot\alpha_2\,{\partial\over\partial\alpha_1}+\cos\alpha_1\,{\partial\over\partial\alpha_2}+{\sin\alpha_1\over\sin\alpha_2}{\partial\over\partial\alpha_3}\cr\cr
G_3&=&\cos\alpha_1\,\cot\alpha_2\,{\partial\over\partial\alpha_1}+\sin\alpha_1\,{\partial\over\partial\alpha_2}-{\cos\alpha_1\over\sin\alpha_2}{\partial\over\partial\alpha_3}\label{su2gensangles}
\eea
It is well known that the eigenfunctions of these Killing vectors are the Wigner functions
\bea
D^l_{m\mu}&=&e^{im\alpha_3}d^l_{m\mu}(\cos\alpha_2)e^{i\mu\alpha_1}
\eea
where
\bea
d^l_{m\mu}(x)&=&\sqrt{(l-m)!(l+m)!\over (l-\mu)!(l+\mu)!}(1-x)^{m+\mu\over 2}(1+x)^{-{m-\mu\over 2}}P^{(-m-\mu,-m+\mu)}_{l+m}(x)
\eea
$P^{(a,b)}_{l+m}(x)$ is a Jacobi polynomial
\bea
P_n^{(a,b)}(x)={(-1)^2\over 2^n n!}(1-x)^{-a}(1+x)^{-b}{d^n\over dx^n}\left[(1-x)^{a+n}(1+x)^{b+n}\right]
\eea
and $l$ runs over the non-negative integers while $m,\mu=-l,-l+1,\cdots,0,1,\cdots,l$. The Wigner functions, evaluated on our angles, can be simplified to
\bea
D^l_{m\mu}&=&\left(\frac{\sqrt{p_3^+}+i \sqrt{p_1^+}}{\sqrt{p_1^++p_3^+}}\right)^m
d^l_{m\mu}\left(\sqrt{p_2^+\over p_1^++p_2^++p_3^+}\right)\cr\cr
&&\quad\times\left(\frac{\sqrt{p_1^+ p_3^+ (p_1^++p_2^++p_3^+)}+i \sqrt{p_2^+} (p_3^+-u_1 (p_1^++p_3^+))}{\sqrt{(p_1^++p_3^+) \left(p_1^+ \left(p_2^+ u_1^2+p_3^+\right)+p_2^+ p_3^+ (u_1-1)^2\right)}}\right)^\mu
\eea
We immediately have ($\mu$ is an eigenvalue while $\kappa$ and $G_1$ are operators)
\bea
G_1 D^l_{m\mu}&=&-\kappa D^l_{m\mu}\,\,=\,\,\mu D^l_{m\mu}
\eea
Thus, the Wigner functions are eigenfunctions of $\kappa$ and hence of the AdS mass operator. They therefore correspond to a definite fall off behaviour of the bulk field as we take $Z\to 0$ as explained in Appendix \ref{boundarykappa}. Two more important properties of the Wigner functions are that they obey an orthogonality relation (i.e. the $D_{m\mu}^l(\alpha_i)$ form a set of orthogonal functions of the angles $\alpha_i$)
\bea
\int _0^{2\pi} d\alpha_3 \int _0^\pi d\alpha_2 \sin \alpha_2 \int _0^{2\pi }d\alpha_1 \,\,D_{m'\mu'}^{l'}(\alpha_1 ,\alpha_2 ,\alpha_3)^{\ast }D_{m\mu'}^{l}(\alpha_1 ,\alpha_2 ,\alpha_3 )={\frac {8\pi ^{2}}{2l+1}}\delta _{m'm}\delta _{\mu'\mu}\delta _{l'l}\cr
&&\label{WFuncs}
\eea
and, by the Peter–Weyl theorem, they form a complete set \cite{WikiWigner}. Consequently, we can use them as a basis in terms of which we can expand the trilocal collective field. It is natural to expect that the coefficients of this harmonic expansion will reproduce the primary operators in the limit that $Z\to 0$. We develop this expectation in the next section.

Finally, we will argue in Section \ref{kangles} that the trilocal field has a natrual interpretation as a field on AdS$_4\times$S$^1\times$S$^2$, where the S$^1$ has coordinate $\alpha_1$ and the $S^2$ has coordinates $\alpha_2,\alpha_3$. One might have expected that the basis provided by $\cos(n\alpha_1)$ and $\sin(n\alpha_1)$, as well as the spherical harmonics  $Y^l_m(\alpha_2,\alpha_3)$ would provide a suitable basis for a harmonic expansion. In addition, this basis seems to be simpler than the basis provided by the Wigner functions. The Wigner functions have a clear advantage. The Wigner functions span irreducible SU(2) representations of the Killing vectors. For a given value of $l$, the Wigner functions $D^l_{m\mu}$ have a definite value for the SU(2) Casimir of the Killing vectors
\bea
C_{2,{\rm SU(2)}}&=&(G_1)^2+(G_2)^2+(G_3)^2\label{su2casop}
\eea
This Casimir has a simple relation to the quadratic Casimir of SO(2,3) ($M,N$ are summed over $-1,0,1,2,3$)
\bea
C_{2,{\rm SO(2,3)}}&=&{1\over 2}L_{MN}L^{NM}\label{quadcas}
\eea
where we write $L_{MN}=-L_{NM}$ in terms of the conformal generators as
\bea
L_{\mu\nu}&=&J_{\mu\nu}\qquad L_{\mu\, -1}\,\,=\,\,{1\over 2}(P_\mu+K_\mu)\cr\cr
L_{-1\,3}&=&-D\qquad\quad L_{\mu 3}\,\,=\,\,{1\over 2}(P_\mu-K_\mu)
\eea
The indices $M$ and $N$ are raised and lowered with the metric $\eta={\rm diag}(-1,-1,1,1,1)$ as usual. The relation between the two Casimirs is
\bea
{1\over\mu(x_i,p^+_i)}C_{2,{\rm SU(2)}}\mu(x_i,p^+_i)&=&-2\left(C_{2,{\rm SO(2,3)}}+\left({3\over 2}\right)^2\right)\label{CasLink}
\eea
where\footnote{The su(2) generators act in the bulk AdS$_4$ theory. The $C_{2,{\rm so(2,3)}}$ Casimir acts in the conformal field theory. The factor $\mu$ defines the usual similarity transformation that must be performed when switching between bulk and boundary. See \cite{deMelloKoch:2024ewt} for more details. The specific form of the conformal generators used are given in Appendix B.1 of \cite{deMelloKoch:2024ewt}.}
\bea
\mu(x_i,p^+_i)&=&\sqrt{p_1^+p_2^+p_3^+}\sqrt{P^+ Z}
\eea
This relation has been determined by direct computation, using mathematica. Thus, the Wigner functions, for a definite value of $l$, span a definite SO(2,3) representation. See Appendix \ref{App1} for further discussion and uses of (\ref{CasLink}).

\subsection{Quadlocal Collective Field}

In Appendix \ref{matchquad} we match the generators of the conformal group for the case of the quadlocal collective field. This allows us to determine the AdS mass operator which plays a central role in the analysis of this section. The AdS mass operator can be written as
\bea
A=\sum_{a=1}^4 \kappa_a^2
\eea
where
\bea
\kappa_d&=&\sum _{a=1}^4 \sum _{b=1}^4 \sum _{c=1}^4 \epsilon_{abcd}
\frac{\sqrt{p^+_a p^+_b}(x_a-x_b) }{2\sqrt{p_c^+}\sqrt{p_1^++p_2^++p_3^++p_4^+}}{\partial\over\partial x_c}
\eea
The $\kappa$ operators close an interesting algebra
\bea
[\kappa_a,\kappa_b]&=&\sum_{c,d=1}^4{\epsilon_{abcd}\sqrt{p^+_c}\over\sqrt{p_1^++p_2^++p_3^++p_4^+}}\kappa_d\label{kappaalagebra}
\eea
These operators are not all independent. They obey the relation
\bea
\sqrt{p_1^+}\kappa_1+\sqrt{p_2^+}\kappa_2+\sqrt{p_3^+}\kappa_3+\sqrt{p_4^+}\kappa_4&=&0
\eea
It is possible to express $M^{XZ}$ in terms of the conformal field theory coordinates as
\bea
M^{XZ}&=&{\sum_{i,j=1}^4 p_j^+(x_i-x_j)^2{\partial\over\partial x_i}-2\sum_{i=1}^4 p_i^+v_i{\partial\over\partial p_i^+}\over 2\sqrt{\sum_{i>j=1}^4 p_i^+p_j^+(x_i-x_j)^2}}
\eea
The AdS mass operator $A$ as well as $M^{XZ}$ should be independent of the bulk AdS coordinates. Using the coordinate transformation defined in (\ref{QuadLocalMap}) we can verify that both the $\kappa_a$ and $M^{XZ}$ can be expressed entirely in terms of the $u_i$ variables. For example, the result for $\kappa_1$ is
\bea
\kappa_1&=&\frac{u_1 u_5 (u_1-u_2)-u_2 u_3+u_3}{\sqrt{u_3 (u_4 u_5+u_4+u_5)+u_4 u_5}}\frac{\partial}{\partial u_1}+\frac{u_2 u_5 (u_1-u_2)+(u_1-1) u_4}{\sqrt{u_3 (u_4 u_5+u_4+u_5)+u_4 u_5}}\frac{\partial}{\partial u_2}
%\cr\cr\cr
%\kappa_2&=&(u_1 u_2 u_5+(u_1-1) (u_2-1)) \sqrt{\frac{u_3}{u_3 (u_4 u_5+u_4+u_5)+u_4 u_5}}\frac{\partial}{\partial u_1}\cr\cr
%&&+\frac{u_3 (u_2 (u_2 u_5+u_2-2)+u_4+1)}{\sqrt{u_3 (u_3 (u_4 u_5+u_4+u_5)+u_4 u_5)}}\frac{\partial}{\partial u_2}\cr\cr\cr
%\kappa_3&=&(-u_1 (u_2 u_5+u_2-1)+u_2-1) \sqrt{\frac{u_4}{u_3 (u_4 u_5+u_4+u_5)+u_4 u_5}}\frac{\partial}{\partial u_2}\cr\cr
%&&-\frac{(u_4 (u_1 (u_1 u_5+u_1-2)+u_3+1))}{\sqrt{u_4 (u_3 (u_4 u_5+u_4+u_5)+u_4 u_5)}}\frac{\partial}{\partial u_1}\cr\cr\cr
%\kappa_4&=&((u_1-1) (u_1-u_2)+u_2 u_3) \sqrt{\frac{u_5}{u_3 (u_4 u_5+u_4+u_5)+u_4 u_5}}\frac{\partial}{\partial u_1}\cr\cr
%&&- ((u_2-1) (u_2-u_1)+u_1 u_4) 
%\sqrt{\frac{u_5}{u_3 (u_4 u_5+u_4+u_5)+u_4 u_5}}
%\frac{\partial}{\partial u_2}
\eea
and $M^{XZ}$ is given by
\bea
M^{XZ}&=&\frac{(u_3 (u_4 u_5+u_4+u_5)+u_4 u_5) }{\sqrt{u_3 u_4 u_5 \left(u_5 (u_1-u_2)^2+u_1 u_4 (u_1 u_5+u_1-2)+u_3 (u_2 (u_2 u_5+u_2-2)+u_4+1)+u_4\right)}}\cr\cr
&&\times\left(u_1 u_3 \frac{\partial}{\partial u_3}+\frac{1}{2} (u_1-1) u_1 \frac{\partial}{\partial u_1}+u_2 u_4 \frac{\partial}{\partial u_4}+\frac{1}{2} (u_2-1) u_2 \frac{\partial}{\partial u_2}+u_5 \frac{\partial}{\partial u_5}\right)
\eea
The $u_i$ coordinates can be used to define five angles. The necessary formulas, which are rather complicated, have been collected in Appendix \ref{angleforquad}. Using these formulas, we find the following form the generators $\kappa_i$ 
\bea
\kappa_1&=&-\cos (\alpha_3){\partial\over\partial\alpha_2}\cr\cr
\kappa_2&=&(\tan (\alpha_1) \sin (\alpha_2) \cos (\alpha_4)+\sin (\alpha_3) \sin (\alpha_4)){\partial\over\partial\alpha_2}+\cos (\alpha_2) \cos (\alpha_4){\partial\over\partial\alpha_1}\cr\cr
\kappa_3&=&(\tan (\alpha_1) (\cos (\alpha_2) \cos (\alpha_5)-\sin (\alpha_2) \sin (\alpha_4) \sin (\alpha_5))+\sin (\alpha_3) \cos (\alpha_4) \sin (\alpha_5)){\partial\over\partial\alpha_2}\cr\cr
&-&(\cos (\alpha_2) \sin (\alpha_4) \sin (\alpha_5)+\sin (\alpha_2) \cos (\alpha_5)){\partial\over\partial\alpha_1}\cr\cr
\kappa_4&=&(\sin (\alpha_3) \cos (\alpha_4) \cos (\alpha_5)-\tan (\alpha_1) (\sin (\alpha_2) \sin (\alpha_4) \cos (\alpha_5)+\cos (\alpha_2) \sin (\alpha_5))){\partial\over\partial\alpha_2}\cr\cr
&+&(\sin (\alpha_2) \sin (\alpha_5)-\cos (\alpha_2) \sin (\alpha_4) \cos (\alpha_5)){\partial\over\partial\alpha_1}
\eea
as well as the map between the original coordinates $x_i$ and $p^+_i$ of the quadlocal collective field and the bulk coordinates is given by
\bea
x_1&=& X+{\cos(\alpha_3)\sin(\alpha_1)\over\sin(\alpha_3)}Z\cr\cr
x_2&=&X-{\sin(\alpha_3)\sin(\alpha_4)\sin(\alpha_1)-\cos(\alpha_4)\cos(\alpha_1)\sin(\alpha_2)\over\cos(\alpha_3)\sin(\alpha_4)}Z\cr\cr
x_3&=&X-{\sin(\alpha_3)\cos(\alpha_4)\sin(\alpha_5)\sin(\alpha_1)+\sin(\alpha_4)\sin(\alpha_5)\cos(\alpha_1)\sin(\alpha_2)-\cos(\alpha_5)\cos(\alpha_1)\cos(\alpha_2)\over\cos(\alpha_3)\cos(\alpha_4)\sin(\alpha_5)}Z\cr\cr
x_4&=&X-{\sin(\alpha_3)\cos(\alpha_4)\cos(\alpha_5)\sin(\alpha_1)+\sin(\alpha_4)\cos(\alpha_5)\cos(\alpha_1)\sin(\alpha_2)+\sin(\alpha_5)\cos(\alpha_1)\cos(\alpha_2)\over\cos(\alpha_3)\cos(\alpha_4)\cos(\alpha_5)}Z\cr\cr
p_1^+&=&P^+ \sin ^2(\alpha_3)\cr\cr
p_2^+&=&P^+ \cos ^2(\alpha_3) \sin ^2(\alpha_4)\cr\cr
p_3^+&=&P^+ \cos ^2(\alpha_3) \cos ^2(\alpha_4) \sin ^2(\alpha_5)\cr\cr
p_4^+&=&P^+ \cos ^2(\alpha_3) \cos ^2(\alpha_4) \cos ^2(\alpha_ 5)
\label{qloccords}
\eea
These are remarkably simple relations, particularly given the awkward intermediate results summarized in Appendix \ref{angleforquad}.

\subsection{$k$-local Collective Field}\label{kangles}

The structure we have found above explains how angles are to be identified in general. There is a geometrical construction that starts with the observation that the mapping between bulk and boundary can be written as
\bea
x_i&=&X+\sqrt{\sum_{l=1}^k p_l^+\over p_i^+}\beta_i Z
\eea
where
\bea
\beta_i&=&{\sqrt{p_i^+}v_i\over Z (\sum_{l=1}^k p_l^+)^{3\over 2}}
\eea
Although it is not manifest from this formula, the $\beta_i$ are functions only of the bulk variables $u_i$ i.e. they are independent of $X$, $Z$ and $P^+$. To see this one must write the right hand side of this last equation entirely in terms of bulk coordinates. The $\beta_i$ define components of a unit vector
\bea
\sum_{i=1}^k\beta_i^2&=&1\label{betasphere}
\eea
Now introduce $k-1$ angles, defined entirely in terms of the light cone momenta as follows
\bea
\tan (\alpha_{k-2+i})&=&\sqrt{p_i^+\over \sum_{l=i+1}^k p_l^+}\qquad\qquad i\,\,=\,\,1,2,\cdots,k-1\label{lastalphas}
\eea
In terms of these angles we now define a set of $k-1$ vectors $\hat{n}_i$, which we call the composite vectors, because they describe how the composite collective field is constructed from the $k$ free scalar fields. These vectors are orthonormal
\bea
\hat{n}_i\cdot\hat{n}_j&=&\delta_{ij}
\eea
They are also all orthogonal to the unit vector $\hat{n}_{p^+}$ which, by definition, has its $i$th entry equal to $\sqrt{p_i^+/P^+}$. The first $i-1$ entries of $\hat{n}_i$ are zero and the $i$th entry is $\cos(\alpha_{k-2+i})$. We then complete the vectors to obtain an orthonormal set. See Appendix \ref{compvec} for examples of the composite vectors for the quadlocal and quintlocal collective fields. From these examples there is an obvious pattern which can be followed to construct the composite vectors for $k$-local collective fields. We now define the final $k-2$ angles with the identification
\bea
\vec{\beta}&=&\sin(\alpha_1)\hat{n}_1+\cos(\alpha_1)\sin(\alpha_2)\hat{n}_2+\cos(\alpha_1)\cos(\alpha_2)\sin(\alpha_3)\hat{n}_3\cr\cr
&&+\cdots+\prod_{i=1}^{k-2}\cos(\alpha_i)\hat{n}_{k-1}\label{parambeta}
\eea
The angles $\alpha_i$ define the complete set of $2k-3$ additional bulk coordinates. Together with $Z,P^+,X$ and $X^+$ this completely accounts for the $2k+1$ coordinates of the $k$-local collective field.

The geometrical structure we have uncovered here is intimately related to the structure of the $\kappa$ algebra (\ref{kappaalagebra}) as we now explain. For concreteness we focus on the case of the quadlocal collective field. However, the argument revolves around the structure of the $\kappa$ operators, and as we explain in Appendix \ref{AdSMssOp}, this has a natural generalization for any $k$-local collective field. For the quadlocal collective field the form of $\beta_i$ is
\bea
\vec{\beta}&=&\sin(\alpha_1)\hat{n}_1+\cos(\alpha_1)\sin(\alpha_2)\hat{n}_2+\cos(\alpha_1)\cos(\alpha_2)\hat{n}_3\label{betaforquad}
\eea
The unit vectors $\vec{n}_1$, $\vec{n}_2$ and $\vec{n}_3$ are functions only of the light cone momenta $p_i^+$ so that
\bea
\kappa_i(\hat{n}_j)&=&0\qquad i=1,2,3,4\quad j=1,2,3
\eea
It is simple to prove that the action of $\kappa$ on $\vec{\beta}$ is given by
\bea
\kappa_i(\beta_j)&=&\sum_{k=1}^4\sum_{l=1}^4\epsilon_{ijkl}(\hat{n}_{p^+})_k\beta_l\label{actforbeta}
\eea
Now, make the ansatz
\bea
\kappa_i=g_i(\alpha_1,\alpha_2){\partial\over\partial\alpha_1}+f_i(\alpha_1,\alpha_2){\partial\over\partial\alpha_2}
\eea
By taking a dot product between $\hat{n}_1$ and $\vec{\beta}$ in (\ref{actforbeta}) we obtain
\bea
\kappa_i(\hat{n}_1\cdot\vec{\beta})=g_i\cos(\alpha_1)=\epsilon_{ijkl}(\hat{n}_1)_j(\hat{n}_{p^+})_k(\cos(\alpha_1)\cos(\alpha_1)
\eea
Next, using the easy to verify result
\bea
\sum_{i=1}^4\sum_{j=1}^4\sum_{k=1}^4\sum_{l=1}^4
\epsilon_{ijkl}(\hat{n}_2)_i (\hat{n}_1)_j (\hat{n}_{p^+})_k(\hat{n}_3)_l=1
\eea
it is a simple exercise to obtain
\bea
\hat{n}_2\cdot\vec{g}&=&\cos(\alpha_2)\qquad\hat{n}_3\cdot\vec{g}\,\,=\,\,-\sin(\alpha_2)\qquad \hat{n}_1\cdot\vec{g}\,\,=\,\,0\,\,=\,\,\hat{n}_{p^+}\cdot\vec{g}
\eea
which implies that
\bea
\vec{g}&=&\cos(\alpha_2)\hat{n}_2-\sin(\alpha_2)\hat{n}_3\,\,=\,\,\left[
\begin{array}{c}
0\\
\cos(\alpha_4)\cos(\alpha_2)\\
-\sin(\alpha_4)\sin(\alpha_5)\cos(\alpha_2)-\cos(\alpha_5)\sin(\alpha_2)\\
-\sin(\alpha_4)\cos(\alpha_5)\cos(\alpha_2)+\sin(\alpha_5)\sin(\alpha_2)
\end{array}
\right]
\eea
This argument has correctly reproduced the coefficient of $\partial_{\alpha_1}$ in the operator $\kappa_i$. A very similar argument shows that
\bea
\vec{f}&=&-\hat{n}_1+\sin(\alpha_2)\tan(\alpha_1)\hat{n}_2+\tan(\alpha_1)\cos(\alpha_2)\hat{n}_3
\eea
which is the correct coefficient of $\partial_{\alpha_2}$ in the operator $\kappa_i$. Finally, note that $\vec{f}\cdot\vec{g}=0$.

It is natural to ask what space the coordinates $\alpha_i$ describe. The statement that $\hat{n}_{p^+}\cdot\hat{n}_{p^+}=1$ is the statement that
\bea
\sum_{i=1}^k (\sqrt{p^+_i})^2=P^+
\eea
This describes a $k-1$ sphere S$^{k-1}$ in the space with coordinates\footnote{Since the $p^+_i$ are all positive they can naturally be interpreted as the square of a coordinate.} $\sqrt{p_i^+}$. The $k-1$ angles defined in (\ref{lastalphas}) are the coordinates of this sphere. See (\ref{trliccoc}) and (\ref{qloccords}) for transparent examples of these angles in the case of the trilocal and quadlocal collective fields. The $\beta_i$ obey $\vec{\beta}\cdot\hat{n}_{p^+}=0$ as well as (\ref{betasphere}). This implies that they describe a $k-2$ sphere S$^{k-2}$ in the space orthogonal to $\hat{n}_{p^+}$. The equation (\ref{parambeta}) describes exactly how the $k-2$ angles $\alpha_i$ $i=1,2,\cdots,k-2$ parametrize this space. Consequently the $2k-3$ angles $\alpha_i$ parametrize the space S$^{k-1}\times$S$^{k-2}$. Thus, the $k$-local collective field is defined on an AdS$_4\times$S$^{k-1}\times$S$^{k-2}$ spacetime and the decomposition of the collective field into bulk fields of a definite spin would exploit harmonic expansions on the product of spheres S$^{k-2}\times$S$^{k-1}$. Based on experience with the trilocal collective field, the correct basis for this expansion is probably not simply spherical harmonics depending on $\alpha_i$ $i=1,2,\cdots k-2$ multiplied by spherical harmonics depending on $\alpha_{k-2+i}$ $i=1,2,\cdots,k-1$. For the trilocal we know that this basis does not respect the SO(2,3) symmetry.

\section{Boundary Condition}\label{BC}

Based on experience\footnote{See formulas (\ref{BiAngle}) and (\ref{spinningcurrent}).} with the bilocal, it is natural to expect that the Wigner functions (\ref{WFuncs}) will reproduce formulas for primary operators when expanded as a power series in the momenta. The analogue of the expansion (\ref{fldident}) is given by
\bea
\Phi&=&\sum_{l,m,\mu}D^l_{m\mu}(\alpha_1,\alpha_2,\alpha_3)H^l_{m,\mu}(X^+,P^+,X,Z)\cr\cr
&=&\sqrt{p_1^+p_2^+p_3^+ P^+ Z}\eta_3(x^+,p_1^+,x_1,p_2^+,x_2,p_3^+,x_3)
\eea
In the same way that the modes $\cos (2s\theta)$ and $\sin (2s\theta)$ appearing in (\ref{fldident}) encode the structure of the two field primary operators (see formula (\ref{BiAngle}) and (\ref{spinningcurrent})), we will see that the Wigner functions $D^l_{m\mu}(\alpha_1,\alpha_2,\alpha_3)$ encode the structure of primary operators constructed using three fields. In order to test this expectation, we begin with a discussion of the relevant primary operators. Primary operators ${\cal O}$ have the lowest dimension in their conformal multiplet so they are annihilated by the special conformal generator $K^\mu$. The primary operators packaged in the trilocal collective field are constructed from three fields. We make the ansatz
\bea
\gamma_s^{++\cdots +}&=&\sum_{k=0}^{s}\sum_{l=0}^{s-k}d_{k,l} (P^+)^k\phi_0 (P^+)^l\phi_0 (P^+)^{s-k-l}\phi_0\label{primaryansatz}
\eea
for these primary operators. Here $\phi_0$ is the free scalar field, which in 2+1 dimensions has dimension ${1\over 2}$. We will determine the coefficients $d_{k,l}$ by requiring that
$K^\mu \gamma_s^{++\cdots +}=0$. The commutators from the complete conformal algebra that we need are
\bea
[K^-,P^+]&=&-D+J^{+-}\qquad\qquad [K^X,P^+]\,\,=\,\,J^{+X}\qquad\qquad [K^+,P^+]=0\cr\cr
[P^+,D]&=&P^+\qquad\qquad [P^+,J^{+X}]\,\,=\,\,0\qquad\qquad [P^+,J^{+-}]=-P^+
\eea
The free scalar field $\phi_0$ obeys
\bea
K^+\phi_0&=&0\qquad\qquad K^-\phi_0\,\,=\,\,0\qquad\qquad K^X\phi_0\,\,=\,\,0\cr\cr
J^{+X}\phi_0&=&0\qquad\qquad J^{+-}\phi_0\,\,=\,\,0\qquad\qquad D\phi_0\,\,=\,\,-{1\over 2}\phi_0
\eea
It is also useful to work out the analogous formulas for the level $k$ descendant of $\phi_0$, given by $(P^+)^k\phi_0$
\bea
K^+ (P^+)^k\phi_0&=&0\qquad K^X (P^+)^k\phi_0\,\,=\,\,0\qquad K^-(P^+)^k\phi_0\,\,=\,\,{(2k-1)k\over 2}(P^+)^{k-1}\phi_0\cr\cr
J^{+X}(P^+)^k\phi_0&=&0\qquad D(P^+)^k\phi_0\,\,=\,\,-(k+{1\over 2})(P^+)^k\phi_0\qquad J^{+-}(P^+)^k\phi_0\,\,=\,\, k (P^+)^k\phi_0\cr
&&
\eea
As a confidence building check of our formulas, note that requiring that $K^\mu \psi_s^{++\cdots +}=0$ with
\bea
\psi_s^{++\cdots +}&=&\sum_{k=0}^{s} c_k (P^+)^k \phi_0 (P^+)^{s-k}\phi_0
\eea
we easily find
\bea
c_k&=&\frac{(-1)^k}{k! (2 k-1)\text{!!} (s-k)! (-2 k+2 s-1)\text{!!}}
\eea
which is the correct answer for the spinning conserved current \cite{Craigie:1983fb}. We easily find
\bea
K^+\gamma_s^{++\cdots +}&=&0\,\,=\,\,K^X\gamma_s^{++\cdots +}
\eea
for any choice of the coefficients $d_{k,l}$, while
\bea
K^-\gamma_s^{++\cdots +}&=&0
\eea
leads to the condition
\bea
\sum_{k=0}^{s-1}\sum_{l=0}^{s-k-1}\tilde{d}_{k,l} (P^+)^k\phi_0 (P^+)^l\phi_0 (P^+)^{s-k-l}\phi_0&=&0\label{vanishing}
\eea
where
\bea
\tilde{d}_{k,l}&=&d_{k+1,l}(2k+1)(k+1)+d_{k,l+1}(2l+1)(l+1)+d_{kl}(2s-2k-2l-1)(s-k-l)\cr
&&\label{tds}
\eea
The condition (\ref{vanishing}), when written in terms of the matrix valued fields, becomes
\bea
\sum_{k=0}^{s-1}\sum_{l=0}^{s-k-1}\tilde{d}_{k,l} \Tr\Big(\partial^{+\,k}\phi\,\partial ^{+\,l}\phi\,\partial^{+\, s-k-l}\phi\Big)&=&0
\eea
Taking the cyclicity of the trace into account, this implies that
\bea
\tilde{d}_{k,l}+\tilde{d}_{l,s-k-l}+\tilde{d}_{s-k-l,k}&=&0\label{ForFirstPC}
\eea
The condition (\ref{ForFirstPC}) must be obeyed by the coefficients $d_{k,l}$ of any primary operator constructed from the trace of derivatives of three scalar fields. We are now in a position to explore the connection between the Wigner functions and primary operators.

To start, consider the Wigner function
\bea
D^l_{00}&=&P_l \left(\sqrt{p_2^+\over p_1^++p_2^++p_3^+}\right)
\eea
with $P_l(x)$ a Legendre polynomial. Using the known expansion of these polynomials we find that
\bea
(p_1^++p_2^++p_3^+)^s D^{2s}_{00}&=&{1\over 2^{2s}}\sum_{k=0}^s (-1)^k {2s \choose k}{4s-2k\choose 2s}(p_2^+)^{s-k}(p_1^++p_2^++p_3^+)^k\cr\cr
&=&\sum_{k=0}^{s}\sum_{l=0}^{s-k}d_{k,l} (p_1^+)^k (p_2^+)^l (p_3^+)^{s-k-l}
\label{onetypeofprimary}
\eea
where
\bea
d_{k,l}&=&\sum _{q=k}^s \frac{(-1)^q (4 s-2 q)! \theta (l+q-s)}{2^{2 s} k! (2 s-q)! (2 s-2 q)! (l+q-s)! (s-l-k)!}
\eea
and our convention for the Heaviside function is that $\theta(x)=1$ for $x\ge 0$. With this formula it is straight forward to check that $\tilde{d}_{k,l}=0$ so that (\ref{ForFirstPC}) is indeed obeyed. This is our first indication that the basis functions of our harmonic expansion, the Wigner functions, do indeed produce primary operators.

Notice that (\ref{onetypeofprimary}), after translating the momenta into derivatives, has produced a primary operator with all indices set to $+$. Working with the equal $x^+$ bilocal provides a description of the dynamics in which the $-$ polarization indices have been eliminated\cite{deMelloKoch:2023ngh}. The $x$ polarizations must still appear, which implies that we should we should obtain a polynomial in $p^+$ (supplying the $\partial_{x^-}$ derivatives) and in $p$ (supplying the $\partial_x$ derivatives). A simple example of a primary operator that does have $x$ polarizations, is obtained by studying a mode with $\kappa=1$. In Appendix \ref{boundarykappa} we argue that any normalizable solution behaves as $Z^{|\kappa|}$ as $Z\to 0$. The primaries we have discussed up to this point have all had $\kappa=0$ so that they tend to a finite non-zero value as we go to the boundary $Z\to 0$. The modes with $\kappa=1$ correspond to fields that vanish as $Z$. To obtain a non-zero boundary value for these fields, we need to take a derivative with respect to $Z$. The derivative with respect to $Z$ has been given in (\ref{derblkcoordinates}). A simple manipulation shows that
\bea
{\partial\over\partial Z}&=&\frac{p_1 (p_2^+ u_1+p_3^+)+p_2 (p_3^+ (1-u_1)-p_1^+ u_1)+p_3 (p_2^+ (u_1-1)-p_1^+)}{\sqrt{p_1^+ p_2^+ u_1^2+p_1^+ p_3^++p_2^+ p_3^+ (1-u_1)^2}}\label{Z}
\eea
Recall that the parameter $u_1$, which tracks which OPE channel we are using, is given by
\bea
u_1={x_1-x_2\over x_1-x_3}
\eea
Consider the channel in which we first take the product of the field at $x_1$ with the field at $x_2$. When taking the first OPE we have $x_1=x_2\ne x_3$. This sets $u_1=0$ and we have
\bea
\left.{\partial\over\partial Z}\right|_{u_1=0}&=&\frac{p_3^+ (p_1+p_2)-p_3 (p_1^++p_2^+)}{\sqrt{p_3^+ (p_1^++p_2^+)}}\label{Z0}
\eea
The momentum operators, for the composite field at $x_1=x_2$, are given by $p_1+p_2$ and $p_1^++p_2^+$. The above formula thus looks rather natural. The appearance of square roots in the momenta looks unnatural. We will see that all square roots cancel out of the final formulas. Next consider the OPE channel in which we first take the product of the field at $x_2$ with the field at $x_3$. In this case case, when taking the first OPE we have $x_3=x_2\ne x_1$. This sets $u_1=1$ and
\bea
\left.{\partial\over\partial Z}\right|_{u_1=1}&=&\frac{p_1 (p_2^++p_3^+)-(p_2+p_3) p_1^+}{\sqrt{p_1^+ (p_2^++p_3^+)}}\label{Z1}
\eea
We see the appearance of the momentum operator for the composite field at $x_2=x_3$, as well as the awkward square root factors. Finally, consider the OPE channel in which we first take the product of the operators at $x_1=x_3$. This time, when taking the first OPE we have $x_1=x_3\ne x_2$ which sets $u_1=\infty$ and\footnote{Recall that there is a $\mathbb{Z}_3$ symmetry as a result of cyclicity of the trace. It is interesting to note that under this $\mathbb{Z}_3$ we have the orbit $u_1\to u_1'=1-u_1^{-1}\to u_1'' =(1-u_1)^{-1}\to u_1$. Consequently, $u_1=\infty$ maps to $u_1'=1$ and then to $u_1''=0$ so that all three channels are related by $\mathbb{Z}_3$.}
\bea
\left.{\partial\over\partial Z}\right|_{u_1=\infty}&=&\frac{p_2^+ (p_1+p_3)-p_2 (p_1^++p_3^+)}{\sqrt{p_2^+ (p_1^++p_3^+)}}\label{Zinf}
\eea
It is this last channel that is relevant for our analysis. In this case, the OPE first produces a composite from the fields at $x_1$ and $x_3$. This matches how the angle $\alpha_2$ and $\alpha_3$ are defined. The angle $\alpha_2$ is defined using all three fields (it is a function of $p^+_i$ for $i=1,2,3$) but $\alpha_3$ is defined using only the fields at $x_1$ and $x_3$. The composite that appears in the $u_1=\infty$ channel is constructed from the fields at $x_1$ and $x_3$. Now consider the Wigner function with $l=4$, $m=0$ and $\mu=1$ which is given by
\bea
D^4_{01}&=&i \sqrt{5} (-3 p_1^++4 p_2^+-3 p_3^+) \sqrt{p_2^+ (p_1^++p_3^+)}\cr\cr
&\times&\frac{ \left(p_2^+ u_1 (p_1^++p_3^+)+i \sqrt{p_1^+ p_2^+ p_3^+ (p_1^++p_2^++p_3^+)}-p_2^+ p_3^+\right)}{4 \sqrt{p_2^+ \left(p_2^+ (p_1^+ u_1+p_3^+ (u_1-1))^2+p_1^+ p_3^+ (p_1^++p_2^++p_3^+)\right)}}
\eea
As we explained above, we need to take a derivative of the mode this field multiplies, with respect to $Z$. We also focus on the OPE channel specified by $u_1=\infty$\footnote{In this channel we have $\alpha_1=-{\pi\over 2}$ so the non-trivial dependence on $\mu$ is in $d^l_{m\mu}$.}. After removing an awkward overall constant we obtain
\bea
&&\left.(p_1^++p_2^++p_3^+)^2\,D^4_{01}\,{\partial\over\partial Z}\right|_{u_1=\infty}\,\,=\,\,p_1 \left((p_2^+)^2-p_1^+ (p_2^++p_3^+)+(p_3^+)^2\right)\cr\cr &&\qquad+p_2\left((p_1^+)^2-p_2^+ (p_1^++p_3^+)+(p_3^+)^2\right)
+p_3 \left((p_1^+)^2-p_3^+ (p_1^++p_2^+)+(p_2^+)^2\right)
\eea
To interpret this mode, consider the energy momentum tensor of the conformal field theory
\bea
T_{\mu\nu}&=&\Tr\left(\partial_\mu\phi\partial_\nu\phi-{\eta_{\mu\nu}\over 2}\partial^\alpha\phi\partial_\alpha\phi\right)+{1\over 8}\Tr\left(\eta_{\mu\nu}\partial_\alpha \partial^\alpha(\phi^2)-\partial_\mu\partial_\nu (\phi^2)\right)
\eea
The second term above is the usual improvement needed to make $T_{\mu\nu}$ traceless. This formula defines a primary operator. Generalizing this formula we can construct the three field primary operator\footnote{This simple generalization is only sure to work in the free field theory where anomalous dimensions are not generated and the generators of the conformal group are simply given by the coproduct with no additional corrections.}
\bea
{\cal O}_{\mu\nu}&=&{1\over 2}\Tr\left(\phi\partial_\mu\phi\partial_\nu\phi+\phi\partial_\nu\phi\partial_\mu\phi-\eta_{\mu\nu}\phi\partial^\alpha\phi \partial_\alpha\phi\right)+{1\over 8}\Tr\left(\eta_{\mu\nu}\phi\partial_\alpha \partial^\alpha(\phi^2)-\phi\partial_\mu\partial_\nu (\phi^2)\right)\nonumber
\eea
From this formula we read off the components
\bea
{\cal O}^{++}&=&-{1\over 4}\left(\phi\phi\partial^{+\, 2}\phi-3\phi\partial^+\phi\partial^+\phi\right)\cr\cr
{\cal O}^{+x}&=&-{1\over 8}\left(2\phi\phi\partial^+\partial^x \phi-3\phi\partial^+\phi\partial^x\phi-3\phi\partial^x\phi\partial^+\phi\right)
\eea
These three field primary operators translate into the following polynomials
\bea
{\cal O}^{++}&=&{1\over 4}(p_1^{+\,2}-3p_1^+p_2^++p_2^{+\, 2}-3p_2^+p_3^++p_3^{+\, 2}-3p_3^+p_1^+)\cr\cr
{\cal O}^{+x}&=&{1\over 8}(2p_1^+p_1-3p_1^+p_2-3p_1p_2^++2p_2p_2^+ -3p_2p_3^+-3p_2^+p_3\cr\cr
&&\quad +2p_3^+p_3-3p_3p_1^+-3p_3^+p_1)
\eea
It is now simple to verify that
\bea
\left.D^4_{01}\,{\partial\over\partial Z}\right|_{u_1=\infty}&=&{8\over 5}(p_1+p_2+p_3){\cal O}^{++}-{8\over 5}(p_1^++p_2^++p_3^+){\cal O}^{+x}\label{ABehaviour}
\eea
The two coefficients, which are sums of momenta, become total derivates when translated back to position space. Consequently, after going back to position space, the $\kappa=1$ bulk mode that we are studying becomes, in the $Z\to 0$ limit, a sum of two descendants
\bea
{8\over 5}\left(\partial^x{\cal O}^{++}-\partial^+{\cal O}^{x+}\right)
\eea
The fact that we only find operators belonging to a particular SO(2,3) representation is a consequence of the fact that our harmonic expansion uses the Wigner functions, as explained in Section \ref{trilocalanalysis}. To connect with that discussion, note that our primaries are built from three fields, with two symmetric and traceless combinations of derivatives so that it has dimension $\Delta={7\over 2}$ and spin $s=2$, leading to
\bea
C_{2,{\rm SO(2,3)}}=\Delta (\Delta-3)+s(s+1)={31\over 4}
\eea
With our conventions, the values of the SU(2) Casimir for the Wigner function $D^l_{m\mu}$ is $-l(l+1)$, which gives $-20$ for our example. Thus,  (\ref{CasLink}) shows that we are matching the correct primary to the Wigner function we study. The behaviour (\ref{ABehaviour}) of the asymptotic bulk field closely matches the behaviour found in \cite{Mintun:2014gua} for the bilocal collective field. There two derivatives of the bulk field, $(\partial^+)^2 h^{xz}$ evaluated at the boundary, reproduces the sum of descendants $\partial_- j_{-x}-2\partial_x j_{--}$ of the primary conserved current. Together these results are compelling evidence that the GKPW rule is reproduced.

The papers \cite{deMelloKoch:2017caf,deMelloKoch:2017dgi,{DeMelloKoch:2018hyq},deMelloKoch:2018klm} have developed methods to construct primary operators in free field theory. Above we have translated operators with derivatives into polynomials in momenta, according to the rule
\bea
\sum_{k,l}c_{k,l}\Tr\Big(\partial^{+\,k}\phi\,\partial ^{+\,l}\phi\,\partial^{+\, s-k-l}\phi\Big)&\to& \sum_{k,l}c_{k,l} (p_1^+)^k(p_2^+)^l(p_3^+)^{s-k-l}
\eea
The papers \cite{deMelloKoch:2017caf,deMelloKoch:2017dgi,{DeMelloKoch:2018hyq},deMelloKoch:2018klm} prove that if this polynomial in the momenta is translated into a polynomial $f(x_1,x_2,x_3)$ by replacing 
\bea
(p_k^+)^i&\to& (2i-1)!!\, (x_k)^i\label{primtopoly}
\eea
then the resulting polynomial $f(x_1,x_2,x_3)$ is invariant under the action of $\mathbb{Z}_3$ on $x_1,x_2,x_3$ (by cyclicity of the trace) and they are harmonic (by the free Klein-Gordon equation). If $f(x_1,x_2,x_3)$ is also translation invariant, then the corresponding operator is primary. Using mathematica it is a straight forward exercise to evaluate the Wigner functions $D^l_{m\mu}$ for a given choice of $l,m,\mu$ and then to use the rule (\ref{primtopoly}) to test if the resulting polynomial is translation invariant. For $\mu\ne 0$ we also need to take $|\mu|$ derivatives with respect to $Z$.

In Appendix \ref{App1} we explicitly identify the Wigner functions associated to the primaries of low dimension. This demonstrates that the primary operators of the conformal field theory are indeed reproduced by the boundary behaviour of the collective field in precisely the manner dictated by the GKPW rule. 

\section{Discussion and Conclusions}\label{conclusions}

In this article we have studied the holography of a free matrix in 2+1 dimensions, using collective field theory. The rearrangement of degree of freedom performed by collective field theory ensures that the original loop expansion parameter of the conformal field theory ($\hbar$) is replaced by loop expansion parameter $1/N$. This is the loop expansion parameter of the dual gravity theory. Collective field theory uses invariant variables as the basic degrees of freedom. In our problem these are traces of a product of matrices, with each evaluated at the same $x^+$ but at distinct $x^-$ and $x$ coordinates. Consequently, the collective field constructed from $k$ matrices is a $k$-local field, depending on $x^+$ and the $x^-_i,x_i$ coordinates of each field. The central result of this study is the interpretation of the complete collection of $2k+1$ coordinates. This is more compelling evidence that collective field theory provides a construction of the AdS/CFT duality.

Our study has made use of recent progress related to understanding bulk locality in the collective description \cite{deMelloKoch:2024juz} and to the description of multilocal operators \cite{deMelloKoch:2024ewt}. Bulk locality dictates how the AdS$_4$ coordinates are constructed from those of the multi-local collective fields. The equations of motion then map into the statement that the bulk field is dual to a given primary and its descendants. Concretely, the bulk equation of motion amounts to setting the quadratic Casimir of SO(2,3) to the value associated with the corresponding primary operator \cite{deMelloKoch:2024ewt}. In this article we have demonstrated that all $2k+1$ coordinates of the collective field have an interpretation in the dual gravity theory: the $2k+1$ coordinates of the $k$-local collective field parametrize the AdS$_4\times$S$^{k-2}\times$S$^{k-1}$ spacetime. The modes defined by a harmonic expansion on S$^{k-2}\times$S$^{k-1}$ are the bulk fields of the dual gravity theory. Further we have demonstrated that the orthogonal basis functions of the harmonic expansion are closely related to primary operators. In particular we have demonstrated that the leading terms of the bulk field in the $Z\to0$ limit reduces to three field primary operators, as dictated by the GKPW rule. This is a very concrete extension of the proposals for holography of the vector model first outlined in the prescient paper \cite{Das:2003vw} and realized in detail in \cite{deMelloKoch:2010wdf}.

A number of directions in which this work can be extended immediately suggest themselves. The fact that internal spheres S$^{k-2}\times$S$^{k-1}$ collect the primary operators packaged into the $k$-local collective field hints at new structures underlying the conformal field theory. Understanding this structure in more detail as well as understanding its origin would be extremely interesting. We have focused only on the trilocal ($k=3$) collective fields. Higher values of $k$ should be studied. The connection between harmonic expansions on spheres and the construction of primary operators that we have uncovered is a very natural extension of the bilocal case studied in \cite{deMelloKoch:2014vnt}. It would be instructive to establish this result for all $k$. Even for $k=3$ important questions remain. We have outlined one interesting question at the end of Appendix \ref{App1}.

Our study has all been carried out for the free field theory. It would be interesting to turn on interactions. In that case too, collective field theory dictates that the dynamics should be written in terms of invariant fields. One must again face the question of the interpretation of the extra coordinates that the collective field depends on. In the free case we studied here, the extra coordinates parametrize internal spheres that collect all the primaries packaged into the invariant field. Does this rather remarkable structure continue unchanged for interacting theories? If not, what replaces it?

Finally, it would be interesting to obtain the holographic mapping relevant for equal time collective fields. This collective field theory description would have a natural extension to finite temperature \cite{Jevicki:2015sla,Jevicki:2017zay,Jevicki:2021ddf} and has the potential to shed light on holography for spacetimes with horizons. 

\begin{center} 
{\bf Acknowledgements}
\end{center}
RdMK is supported by a start up research fund of Huzhou University, a Zhejiang Province talent award and by a Changjiang Scholar award. PR is also supported by the South African Research Chairs Initiative of the Department of Science and Technology and the National Research Foundation. HJRVZ is supported in part by the “Quantum Technologies for Sustainable Development” grant from the National Institute for Theoretical and Computational Sciences of South Africa (NITHECS).

\begin{appendix}

\section{Primary polynomials}\label{App1}

Following the papers \cite{deMelloKoch:2017caf,deMelloKoch:2017dgi,{DeMelloKoch:2018hyq},deMelloKoch:2018klm}, polynomials corresponding to the primary operators packaged in the trilocal collective field were written down in \cite{deMelloKoch:2024ewt}. In this Appendix we will identify some of these polynomials in terms of Wigner functions. This gives additional evidence that we do indeed reproduce the primary operators as $Z\to 0$ in the collective field theory description.

According to \cite{deMelloKoch:2024ewt} the polynomials corresponding to primary operators that can be obtained from the trilocal collective field are given by 
\bea
g_1^n\, g_2^m\,\qquad n,m=0,1,2,\cdots
\label{gg}
\eea
\bea
g_1^n\, g_2^m\, \epsilon\cdot v\qquad n,m=0,1,2,\cdots
\label{ggv}
\eea
\bea
g_1^n\, g_2^m\, Q_1\qquad n,m=0,1,2,\cdots
\label{ggq1}
\eea
and 
\bea
g_1^n\, g_2^m\, Q_2\qquad n,m=0,1,2,\cdots
\label{ggq2}
\eea
where the building blocks for these polynomials are
\bea
g_1&=&(\epsilon\cdot (x_1-x_2))^2+(\epsilon\cdot (x_2-x_3))^2+(\epsilon\cdot (x_3-x_1))^2\cr\cr
g_2&=&\epsilon\cdot (x_1+x_2-2x_3)\epsilon\cdot (x_2+x_3-2x_1)
\epsilon\cdot (x_3+x_1-2x_2)\label{HSgenerators}
\eea
as well as
\bea
v^{\rho}&=&\epsilon^{\mu\nu\rho}\left(x_{1,\mu}x_{2,\nu}+x_{2\mu}x_{3\nu}+x_{3\mu}x_{1\nu}\right)\label{vdefn}
\eea
\bea
Q_1&=&\epsilon\cdot(x_1-x_2)\,\epsilon\cdot(x_2-x_3)\,\epsilon\cdot(x_3-x_1)
\eea
\bea
Q_2&=&\epsilon\cdot(x_1-x_2)\,\epsilon\cdot(x_2-x_3)\,\epsilon\cdot(x_3-x_1)\, \epsilon_\rho\epsilon^{\mu \nu \rho}  \left( x_{1,\mu} x_{2,\nu} + x_{2,\mu} x_{3,\nu} + x_{3,\mu} x_{1,\nu}    \right)
\eea
These should all be expressible in terms of Wigner functions. For low degrees we have explicitly demonstrated that this is indeed the case. Here are some examples
\bea
g_1&=&{8\over 21}{\cal P}\left( (p_1^++p_2^++p_3^+)^2D^4_{00}\right)\cr\cr
g_2&=&-{16\over 45}{\cal P}\left( (p_1^++p_2^++p_3^+)^3 D^6_{00}\right)\cr\cr
Q_1&=&-{16\over 63\sqrt{105}}{\cal P}\left( (p_1^++p_2^++p_3^+)^3 (D^6_{20}+D^6_{-20})\right)\cr\cr
g_1^2&=&{256\over 10395}{\cal P}\left( (p_1^++p_2^++p_3^+)^4 D^8_{00}\right)\cr\cr
g_1g_2&=&-{512\over 61425}{\cal P}\left( (p_1^++p_2^++p_3^+)^5 D^{10}_{00}\right)\cr\cr
g_1Q_1&=&-{256\over 77805}\sqrt{2\over 165}{\cal P}\left( (p_1^++p_2^++p_3^+)^5 (D^{10}_{20}+D^{10}_{-20})\right)
\eea
where ${\cal P}(\cdot)$ denotes the map from polynomials in momenta ($p_i^+$) to polynomials in $x_i$ by employing the rule (\ref{primtopoly}). The primary polynomials reproduced above have not involved the $v^\rho$ or $Q_2$ structures. These both involve $\epsilon_\rho\epsilon^{\mu \nu \rho}$ which generates anti-symmetric combinations of $p_i$ and $p_i^+$. We have seen that these are introduced by derivatives with respect to $Z$ (see formulas (\ref{Z}), (\ref{Z0}), (\ref{Z1}) and (\ref{Zinf})), so that these primaries will be coded into Wigner functions with $\kappa\ne 0$.

To end this Appendix, we use (\ref{CasLink}) to construct a dictionary between the Wigner functions and the complete set of three field primary operators. By using character methods 
\cite{deMelloKoch:2024ewt} determined the spectrum of primary operators constructed using three fields. The computation evaluates the tensor product of three copies of the representation of the free scalar field, while imposing the $\mathbb{Z}_3$ symmetry which arises from cyclicity of the trace. The result is
\bea
{\rm Cyc}\left(({1\over 2},0)^{\otimes 3}\right)&=& ({3\over 2},0)\oplus \bigoplus_{n=0}^\infty \left( (n+2)\left( {9\over 2} + 3n, 3n+3   \right) \oplus n\left( {9\over 2} + 3n, 3n+2   \right)\oplus   \right.   \cr
& & \oplus (n+1)\left( {9\over 2} + 3n-1, 3n+2   \right) \oplus (n+1)\left( {9\over 2} + 3n-1, 3n+1   \right) \oplus     \cr
& & \left.   \oplus (n+1)\left( {9\over 2} + 3n+1, 3n+4   \right) \oplus (n+1)\left( {9\over 2} + 3n+1, 3n+3   \right)    \right)   \label{coolresult}\cr\cr
&&
\eea
There are two types of primaries appearing on this list. We have primaries with dimension $\Delta ={3\over 2}+s$ where $s$ the spin takes the values $s=0$ or $s=2,3,4,\cdots$. The quadratic Casimir of these primaries is given by
\bea
C_{2,{\rm SO(2,3)}}&=&\Delta(\Delta-3)+s(s+1)\,\,=\,\,2s^2+s-{9\over 4}
\eea
Using the relation (\ref{CasLink}) this corresponds to
\bea
C_{2,{\rm SU(2)}}&=&-\,2s(2s+1)
\eea
We also have primaries with dimension $\Delta ={3\over 2}+s+1$ where the spin now takes values $s=1,2,3,\cdots$. The quadratic Casimir of these primaries is given by 
\bea
C_{2,{\rm SO(2,3)}}&=&2s^2+3s-{5\over 4}
\eea
Again using the relation (\ref{CasLink}) this second family of primaries correspond to
\bea
C_{2,{\rm SU(2)}}&=&-\,(2s+1)(2s+2)
\eea
Now, using the explicit form of the su(2) generators (\ref{su2gensangles}), the definition of the Casimir (\ref{su2casop}) and the known Wigner functions, we easily find
\bea
C_{2,{\rm SU(2)}} D^l_{m\mu}=-l(l+1) D^l_{m\mu}
\eea
Consequently, the primaries with dimension $\Delta={3\over 2}+s$ correspond to bulk fields associated with the Wigner functions $D^{2s}_{m\mu}$ while the primaries with dimension $\Delta={3\over 2}+s+1$ correspond to bulk fields associated with the Wigner functions $D^{2s+1}_{m\mu}$.

This dictionary can be made more precise: each bulk field has a definite spin $s$. A field with spin $s$ is a totally symmetric and traceless rank-$s$ tensor in AdS$_4$. The number of independent components of this field is given by 
\bea
{(s+1)(s+2)(s+3)\over 6}-{(s-1)s(s+1)\over 6}=(s+1)^2
\eea
The bulk field dual to primary operators with dimension $\Delta={3\over 2}+s$ is associated with the Wigner function $D^{2s}_{m\mu}$. Since $m$ and $\mu$ each take a total of $4s+1$ values, there are a total of $(4s+1)^2$ distinct orthogonal functions collected in $D^{2s}_{m\mu}$. Clearly then, it is only a subset of the functions defined by $D^{2s}_{m\mu}$ that are used in defining the bulk fields. Recall that cyclicity of the trace implies there is a $\mathbb{Z}_3$ symmetry that must be imposed. We should impose this symmetry to determine which Wigner functions are admissible\footnote{This is again parallel to the bilocal collective field treatment: in that case the bilocal is invariant under a $\mathbb{Z}_2$ symmetry and this reduces to even spins for the bulk gauge fields.}. Although we have described a number of explicit examples of how primary operators correspond to specific $D^{2s}_{m\mu}$, we have not derived the general rule for the correspondence.

\section{Conformal Transformations for the Quadlocal Collective Field}\label{matchquad}

The bulk higher spin fields are collected into a single field
\bea
\Phi(X^+,X^-,X,Z,\alpha^I)&=&\sum_{s=0}^\infty\alpha_{I_1}\alpha_{I_2}\cdots \alpha_{I_{2s}}{A^{I_1 I_2\cdots I_{2s}}(X^+,X^-,X,Z)\over Z}|0\rangle
\eea
The index $I$ on the oscillators runs over $Z$ and $X$. The quadlocal collective field can be decomposed as\footnote{The coefficient of $\eta_4$ in the next formula is chosen to ensure it has a two point function of order 1 as $N\to\infty$.}
\bea
\sigma_4=\sigma_4^0+{1\over N^2}\eta_4
\eea
where $\sigma_4^0$ is the large $N$ expectation value of $\sigma_4$ and $\eta_4$ is a fluctuation about this large $N$ value. It is $\eta_4$ that is mapped to the bulk higher spin gravity field. The holographic mapping between the fields in this case is
\bea
\Phi(X^+,P^+,X,Z,\alpha^I)&=&\mu(p_i^+,x_i)\eta_4(x^+,p_1^+,x_1,p_2^+,x_2,p_3^+,x_3,p_4^+,x_4)\label{matchfields}
\eea
where $\mu(p_i^+,x_i)$ is a factor needed to ensure that the conformal generators of the collective field theory and those of the higher spin gravity map into each other under the change of spacetime coordinates given in (\ref{QuadLocalMap}). Using $L_{\rm AdS}$ to denote a bulk generator and $L_{\rm CFT}$ to denote the corresponding collective generator, we have
\bea
L_{\rm AdS}=\mu(p_i^+,x_i) L_{\rm CFT}{1\over\mu(p_i^+,x_i)}\label{LeqL}
\eea
By matching generators we learn that
\bea
\mu(p_i^+,x_i)=\sqrt{p_1^+ p_2^+ p_3^+p_4^+}\,P^+Z
\eea
with the expression for $Z$ and $P^+$ given in (\ref{coordmap}) with $k=4$. The formula for the AdS mass operator $A$ is derived matching the $P^-$ generator, while the formula for $M^{XZ}$ is obtained by matching $K^X$ at $x^+=0$. These then determine the $M^{-X}$ generator \cite{Metsaev:1999ui} by the formula
\bea
M^{-X}&=&{1\over P^+}M^{XZ}{\partial\over\partial Z}+{1\over 2ZP^+}[M^{XZ},A]
\eea
The expression for $K^-$ also features the operator $B$ which is given by
\bea
B=-{1\over 2}[M^{XZ},[M^{XZ},A]]+M^{XZ}M^{XZ}
\eea
For completeness we list the generators we used to verify (\ref{LeqL}). The generators $L_{\rm CFT}$ are given by
\begin{eqnarray}
P^+&=&\sum_{i=1}^4 p_i^+\cr\cr
P^x&=&\sum_{i=1}^4{\partial\over \partial x_i}\cr\cr
P^-&=&-\sum_{i=1}^4 {1\over 2p_i^+}{\partial^2\over\partial x_i^2}\cr\cr
J^{+-}&=&x^+ P^- +\sum_{i=1}^4 {\partial\over\partial p_i^+}\, p_i^+\cr\cr
J^{+x}&=&x^+ \sum_{i=1}^4{\partial\over\partial x_i}-\sum_{i=1}^4 x_i p_i^+\cr\cr
J^{-x}&=&\sum_{i=1}^4\left(-{\partial\over\partial p_i^+}{\partial\over\partial x_i}+{x_i\over 2p_i^+}{\partial^2\over\partial x_i^2}\right)\cr\cr
D&=&x^+P^- +\sum_{i=1}^4\left(-{\partial\over\partial p_i^+}p_i^+
+x_i {\partial\over\partial x_i}\right)+2\cr\cr
K^+&=&-{1\over 2}\sum_{i=1}^4\left(-2 x^+ {\partial\over\partial p_i^+}p_i^+
+x_i^2 p_i^+\right)+x^+D\cr\cr
K^-&=& \sum_{i=1}^4\Bigg({3\over 2}{\partial\over\partial p_i^+}
+p_i^+ {\partial^2\over\partial p_i^{+\,\,2}}
-x_i{\partial\over\partial x_i}{\partial\over\partial p_i^+} 
+{x_i^2\over 4p_i^+}{\partial^2\over\partial x_i^2}\Bigg)\cr\cr
K^x&=&-{1\over 2}\sum_{i=1}^4\left(-2 x^+ {\partial\over\partial x_i}{\partial\over\partial p_i^+}+x_i^2 {\partial\over\partial x_i}\right)\cr
&&+\sum_{i=1}^4 x_i \left(-x^+ {1\over 2p_i^+}{\partial^2\over\partial x_i^2}
-{\partial\over\partial p_i^+}p_i^+ + x_i{\partial\over\partial x_i}+{1\over 2}\right)
\label{METGEN}
\end{eqnarray}
The generators $L_{\rm AdS}$ are from \cite{Metsaev:1999ui}. They are given by
\bea
P^X&=&\partial_X\cr\cr
P^+&=&P^+\cr\cr
P^-&=&-{\partial_X^2+\partial_Z^2\over 2P^+}+{1\over 2Z^2P^+}A\cr\cr
D&=&X^+ P^- -P^+\partial_{P^+}+X\partial_X+Z\partial_Z\cr\cr
J^{+-}&=&X^+P^-+P^+\partial_{P^+}+1\cr\cr
J^{+X}&=&X^+\partial_X-XP^+\cr\cr
J^{-X}&=&-\partial_{P^+}\partial_X-XP^-+M^{-X}\cr\cr
K^+&=&-{1\over 2}(X^2+Z^2-2X^+\partial_{P^+})P^++X^+ D\cr\cr
K^X&=&-{1\over 2}(X^2+Z^2-2X^+\partial_{P^+})\partial_X+XD+M^{XZ}Z+M^{X-}X^+\cr\cr
K^-&=&-{1\over 2}(X^2+Z^2-2X^+\partial_{P^+})P^--\partial_{P^+}D+{1\over P^+}(X\partial_Z-Z\partial_X)M^{XZ}\cr\cr
&&-{X\over 2ZP^+}[M^{ZX},A]+{1\over P^+}B
\eea

\section{Angles for the Quadlocal Collective Field}\label{angleforquad}

Define five angles by the relations
\bea
\cot (\alpha_3) \sin (\alpha_1)&=&-\frac{(u_1u_4 u_5+u_3 (u_2u_5+u_4))}{\sqrt{f(u_1,u_2,u_3,u_4,u_5)}}\cr\cr
{\cot(\alpha_4)\cos(\alpha_1)\sin(\alpha_2)-\sin(\alpha_3)\sin(\alpha_1)\over\cos\alpha_3}&=&\frac{u_3(u_5 (u_1-u_2)+u_4 (u_1u_5+u_1-1))}{\sqrt{f(u_1,u_2,u_3,u_4,u_5)}}\cr\cr
\tan (\alpha_3)&=&\sqrt{\frac{u_3 u_4 u_5}{u_3 u_4+u_3 u_5+u_4 u_5}}\cr\cr
\tan (\alpha_4)&=&\sqrt{\frac{u_4 u_5}{u_3 u_4+u_3 u_5}}\cr\cr
\tan (\alpha_5)&=&\sqrt{\frac{u_5}{u_4}}
\eea
where the polynomial $f(u_1,u_2,u_3,u_4,u_5)$ is defined in equation (\ref{defpolyf}). This transformation is invertible with the result
\bea
u_1&=&\frac{\cot (\alpha_4) (\sin (\alpha_1) \csc (\alpha_3)-\cos (\alpha_1) \sin (\alpha_2) \cot (\alpha_4))}{\cos (\alpha_1) (\cos (\alpha_2) \csc (\alpha_4) \tan (\alpha_5)+\sin (\alpha_2))+\sin (\alpha_1) \csc (\alpha_3) \cot (\alpha_4)}\cr\cr
u_2&=&\frac{\cot (\alpha_5) (\cos (\alpha_1) (\sin (\alpha_2) \tan (\alpha_4)-\cos (\alpha_2) \sec (\alpha_4) \cot (\alpha_5))+\sin (\alpha_1) \csc (\alpha_3))}{\cos (\alpha_1) (\sin (\alpha_2) \tan (\alpha_4) \cot (\alpha_5)+\cos (\alpha_2) \sec (\alpha_4))+\sin (\alpha_1) \csc (\alpha_3) \cot (\alpha_5)}\cr\cr
u_3&=& \tan ^2(\alpha_3) \csc ^2(\alpha_4)\cr\cr
u_4&=& \tan ^2(\alpha_3) \sec ^2(\alpha_4) \csc ^2(\alpha_5)\cr\cr
u_5&=& \tan ^2(\alpha_3) \sec ^2(\alpha_4) \sec ^2(\alpha_5)
\eea

\section{Composite Vectors for Quadlocal and Quintlocal Collective Fields}\label{compvec}

In this section we simply list the composite vectors for the quadlocal and quintlocal collective fields. See Section \ref{kangles} for further explanation.

\subsection{Composite vectors for Quadlocal Collective Fields}

The angles below are obtained by setting $k=4$ and then making use of (\ref{lastalphas}). 
\bea
\hat{n}_1&=&\left[
\begin{array}{c}
\cos(\alpha_3)\\
-\sin(\alpha_3)\sin(\alpha_4)\\
-\sin(\alpha_3)\cos(\alpha_4)\sin(\alpha_5)\\
-\sin(\alpha_3)\cos(\alpha_4)\cos(\alpha_5)
\end{array}\right]
\qquad
\hat{n}_2\,\,=\,\,\left[
\begin{array}{c}
0\\
\cos(\alpha_4)\\
-\sin(\alpha_4)\sin(\alpha_5)\\
-\sin(\alpha_4)\cos(\alpha_5)
\end{array}\right]
\qquad
\hat{n}_3\,\,=\,\,\left[
\begin{array}{c}
0\\
0\\
\cos(\alpha_5)\\
-\sin(\alpha_5)
\end{array}\right]\cr\cr
&&
\eea
Note that these vectors are all orthogonal to
\bea
\hat{n}_{p^+}&=&\left[
\begin{array}{c}
\sin(\alpha_3)\\
\cos(\alpha_3)\sin(\alpha_4)\\
\cos(\alpha_3)\cos(\alpha_4)\sin(\alpha_5)\\
\cos(\alpha_3)\cos(\alpha_4)\cos(\alpha_5)
\end{array}\right]
\eea
Thus, these vectors furnish an orthonormal basis. The vectors $\hat{n}_1$ and $\hat{n}_{p^+}$ are functions of the lightcone momenta of all 4 particles. The vector $\hat{n}_2$ depends only on the lightcone momenta of particles 2, 3 and 4. Finally, the vector $\hat{n}_3$ depends only on the lightcone momenta of particles 3 and 4. This summarizes which channel we should use for the OPE when taking the bulk field to the boundary: we first take the product of the fields located at $x_3$ and $x_4$. We then take the product of this result with the field located at $x_2$. We then take the product of this result with the field located at $x_1$. This corresponds to setting $u_1=1=u_2$.

\subsection{Composite vectors for Quintlocal Collective Fields}

The angles below are obtained by setting $k=5$ and then making use of (\ref{lastalphas}). 
\bea
\hat{n}_1&=&\left[
\begin{array}{c}
\cos(\alpha_4)\\
-\sin(\alpha_4)\sin(\alpha_5)\\
-\sin(\alpha_4)\cos(\alpha_5)\sin(\alpha_6)\\
-\sin(\alpha_4)\cos(\alpha_5)\cos(\alpha_6)\sin(\alpha_7)\\
-\sin(\alpha_4)\cos(\alpha_5)\cos(\alpha_6)\cos(\alpha_7)
\end{array}\right]\cr\cr\cr\cr
\hat{n}_2&=&\left[
\begin{array}{c}
0\\
\cos(\alpha_5)\\
-\sin(\alpha_5)\sin(\alpha_6)\\
-\sin(\alpha_5)\cos(\alpha_6)\sin(\alpha_7)\\
-\sin(\alpha_5)\cos(\alpha_6)\cos(\alpha_7)
\end{array}\right]
\qquad
\hat{n}_3\,\,=\,\,\left[
\begin{array}{c}
0\\
0\\
\cos(\alpha_6)\\
-\sin(\alpha_6)\sin(\alpha_7)\\
-\sin(\alpha_6)\cos(\alpha_7)
\end{array}\right]
\qquad
\hat{n}_4\,\,=\,\,\left[
\begin{array}{c}
0\\
0\\
0\\
\cos(\alpha_7)\\
-\sin(\alpha_7)
\end{array}\right]\cr\cr
&&
\eea
Note that these vectors are all orthogonal to
\bea
\hat{n}_{p^+}&=&\left[
\begin{array}{c}
\sin(\alpha_4)\\
\cos(\alpha_4)\sin(\alpha_5)\\
\cos(\alpha_4)\cos(\alpha_5)\sin(\alpha_6)\\
\cos(\alpha_4)\cos(\alpha_5)\cos(\alpha_6)\sin(\alpha_7)\\
\cos(\alpha_4)\cos(\alpha_5)\cos(\alpha_6)\cos(\alpha_7)
\end{array}\right]
\eea
Thus, these vectors again furnish an orthonormal basis. The angle $\alpha_7$ depends only on $p_4^+$ and $p_5^+$. The angle $\alpha_6$ depends only on $p_3^+$, $p_4^+$ and $p_5^+$. The angle $\alpha_5$ depends only on $p_2^+$, $p_3^+$, $p_4^+$ and $p_5^+$. Finally the angle $\alpha_4$ depends on all of the $P_i^+$. This again summarizes which channel we should use for the OPE when taking the bulk field to the boundary: we first take the product of the fields located at $x_4$ and $x_5$. We then take the product of this result with the field located at $x_3$. We then take the product of this result with the field located at $x_2$. Finally, we take the product of this result with the field located at $x_1$. This corresponds to setting $u_1=u_2=u_3=1$.

\section{Comment on the AdS mass operator}\label{AdSMssOp}

By matching the generators, using (\ref{LeqL}) as before, acting on the quintic collective field with those of the dual gravity given in (\ref{METGEN}), we learn that
\bea
\mu(p_i^+,x_i)&=&\sqrt{p_1^+ p_2^+ p_3^+p_4^+p^+_5}\,(P^+Z)^{3\over 2}
\eea
Further we learn that the AdS mass operator can be written as
\bea
A&=&{1\over 2}\sum_{i=1}^5\sum_{j=1}^5\kappa_{ij}\kappa_{ij}-{3\over 4}
\eea
where
\bea
\kappa_{ij}&=&\sum_{k=1}^5\sum_{l=1}^5\sum_{m=1}^5\epsilon_{ijklm}{\sqrt{p_k^+p_l^+}(x_k-x_l)\over 2\sqrt{p_m^+(p_1^++p_2^++p_3^++p_4^++p_5^+)}}{\partial\over\partial x_m}
\eea
These results together with those of the trilocal and quadlocal collective fields lead us to conjecture that for the $k$-local collective field we have
\bea
\mu(p_i^+,x_i)&=&\sqrt{p_1^+ p_2^+ p_3^+p_4^+\cdots p^+_k}\,(P^+Z)^{k-2\over 2}
\eea
It is also natural to conjecture that the AdS mass operator is given by
\bea
A&=&{1\over (k-3)!}\sum_{i_1=1}^k\sum_{i_2=1}^k\cdots \sum_{i_{k-3}=1}^k
\kappa_{i_1 i_2\cdots i_{k-3}}\kappa_{i_1 i_2\cdots i_{k-3}}+{\rm constant}
\eea
where
\bea
\kappa_{i_1 i_2\cdots i_{k-3}}&=&\sum_{i_{k-2}=1}^k\sum_{i_{k-1}=1}^k\sum_{i_k=1}^k\epsilon_{i_1i_2\cdots i_{k-1}i_k}{\sqrt{p^+_{k-2}p^+_{k-1}}(x_{k-2}-x_{k-1})\over 2\sqrt{p_k^+(p_1^++p_2^++\cdots+p_k^+)}}{\partial\over\partial x_k}
\eea
The form of these operators plays an important role (as explained in Section \ref{kangles}) in exhibiting the geometrical structure associated to the extra angle coordinates. The fact that there is a nice angle formula for any $k$ is intimately related to the fact that the $\kappa$ operators defined in this Appendix have a nice generalization.

\section{Boundary behaviour of bulk fields}\label{boundarykappa}

In this section we would like to determine the behaviour of the trilocal collective field as $Z\to 0$, from the behaviour of the dual bulk fields. Recall that the equations of motion for the bulk fields are given by \cite{Metsaev:1999ui,Metsaev:2003cu,Metsaev:2005ws,Metsaev:2013kaa,Metsaev:2015rda,Metsaev:2022ndg}
\bea
\left(2\partial^+\partial^-+\partial_X^2+\partial_Z^2-{A\over Z^2}\right)\Phi(X^+,P^+,X,Z,\alpha^I) &=&0\label{blkeqnmot}
\eea
where the AdS mass operator $A$ can be written as 
\bea
A=\kappa^2-{1\over 4}
\eea
Assuming that the bulk field $\Phi(X^+,P^+,X,Z,\alpha^I)$ behaves as $Z^\alpha$ as $Z\to 0$, the most singular terms of (\ref{blkeqnmot}) imply that
\bea
\alpha (\alpha-1)-A&=&0\qquad\qquad\Rightarrow\quad\alpha={1\over 2}\pm \kappa
\eea
Now, taking into account the relation between the trilocal collective field and the bulk gravity field
\bea
\Phi(X^+,P^+,X,Z,\alpha^I)&=&\sqrt{p_1^+p_2^+p_3^+}\sqrt{ZP^+}\eta_3(x^+,p_1^+,x_1,p_2^+,x_2,p_3^+,x_3)\label{matchfields}
\eea
we learn that $\eta_3$ behaves as $Z^{\pm\kappa}$ as $Z\to 0$ so that $\kappa$ controls the boundary behaviour of the different bulk fields packaged in the collective field. Since we are interested in the normalizable solutions, it is clear that after translating  to bulk coordinates, the component of $\eta_3$ corresponding to eigenvalue $\kappa$ vanishes as $Z^\kappa$. To extract this component from the collective field we should take $\kappa$ derivatives with respect to $Z$ before taking $Z\to 0$.

\end{appendix}

\end{document}